\documentclass[aps,prb,twocolumn,groupedaddress,floats,showpacs]{revtex4}

\usepackage{bm}
\usepackage{amssymb,amsmath}
\usepackage{latexsym}
\usepackage{graphicx}
\usepackage{epsfig}

\begin{document}

\title{
Manipulating   Femtosecond Spin--Orbit Torques 
with  Laser Pulse Sequences  to 
Control  
Magnetic Memory States and  Ringing}
\author{P. C. Lingos$^{1}$}
\author{J. Wang$^3$}
\author{I. E. Perakis$^{1,2}$}
\affiliation{$^{1}$Department of Physics, University of Crete,
 Box 2208, Heraklion, Crete, 71003, Greece}
\affiliation{$^{2}$Institute of Electronic Structure 
\& Laser, Foundation for
Research and Technology-Hellas, Heraklion, Crete, 71110, Greece}
\affiliation{$^{3}$Ames Laboratory and Department of 
Physics and Astronomy, Iowa State
University, Ames, Iowa 50010, USA}
\date{\today}

\begin{abstract}
Femtosecond (fs) coherent control of collective
order parameters
is  important for
non--equilibrium  phase dynamics
in  correlated materials.
Here we propose   a  possible scheme for
fs   control
of a ferromagnetic order parameter
based on non--adiabatic
optical manipulation of
 electron--hole ($e$--$h$)
photoexcitations
between spin--orbit--coupled
 bands that are
exchange--split by  magnetic interaction
with
local spins.
We photoexcite
fs carrier spin--pulses
with controllable direction and  time profile
 without using circularly--polarized  light,
via
time--reversal symmetry--breaking
by non--perturbative
interplay between spin--orbit and magnetic exchange
coupling of coherent photocarriers.
We manipulate
photoexcited
{\em fs  spin--orbit torques} to
control complex
switching pathways of the magnetization
between multiple magnetic memory states.
We calculate the photoinduced fs  magnetic anisotropy
in the time domain by using density matrix equations of motion
rather than
the quasi--equilibrium free energy.
By comparing to  pump--probe experiments,
we identify a
``sudden''  magnetization canting induced by
laser excitation,  which
displays magnetic hysteresis
absent in static magneto--optical measurements
and agrees with switchings measured by  Hall magnetoresistivity.
The  fs
 magnetization canting
switches direction with  magnetic state
and  laser frequency, which
distinguishes it  from  nonlinear optical
and demagnetization longitudinal
effects.
By
shaping two--color
laser--pulse
sequences
 analogous to
multi--dimensional Nuclear Magnetic Resonance (NMR)
 spectroscopy,
we show
that
sequences
of clockwise or  counter--clockwise
fs spin--orbit torques
can enhance or suppress
 magnetic ringing and switching
rotation at any desired time.
We propose protocols
that can provide
controlled access to  four
 magnetic states via consequative  90$^{o}$
 switchings.
 \end{abstract}


\maketitle

\section{Introduction}

Femtosecond (fs)  control
 of  switching between condensed matter states 
\cite{li-2013,
Li-2,kapetanakis-2009,
jong}
 may 
address challenges posed by 
multi--functional 
 devices for  information 
storage and processing on a single chip  at 
up--to--thousand--times faster terahertz speeds.
One of the main 
obstacles for  widespread use of magnetic materials 
in such applications 
is the lack of efficient control of magnetization.
Fast spin manipulation is one of the main challenges 
for spin–-electronics, spin--photonics, magnetic storage, 
and quantum computation. \cite{fert} 
To meet this challenge,  different magnetic systems must 
be explored.
In 
diverse  systems
ranging  from  ferromagnetic semiconductors  
\cite{jung,burch,wang-rev}
to doped topological insulators, 
\cite{TI,TI2} 
magnetic effects 
arise from 
 exchange 
interactions ($\propto{\bf S}\cdot{\bf s}$)
between two distinct 
sub--systems: 
mobile,   spin--orbit--coupled 
electron 
 spins (${\bf s}$)
and 
magnetic local  moments (${\bf S}$).\cite{nagaev} 
These interactions couple,
for example,  
  magnetic impurity spins   with
Dirac fermions in topological insulators \cite{TI} or  
 valence--band holes in (III,Mn)V semiconductors.\cite{jung}
Such  couplings
 break time--reversal symmetry 
and resut in 
 ferromagnetic states 
with two 
distinct but strongly--coupled
collective--spin order parameter components. \cite{jung,TI}
When brought out of 
thermodynamic equilibrium,  
 interacting 
mobile 
 and local collective spins 
allow  more ``knobs'' 
for manipulating 
ultrafast magnetism
\cite{bigot1} by using 
 fs laser pulses. 

\begin{figure*} 
\begin{center} 
\includegraphics[scale=0.55]{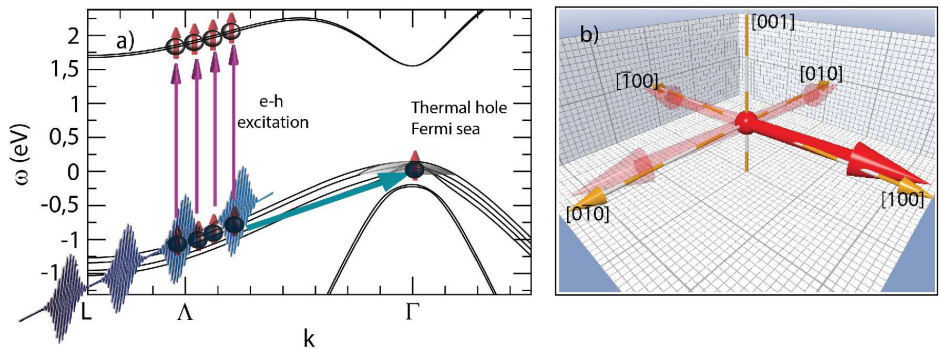} 
\caption{(Color online)
(a): 
Schematic of 
two contributions to the 
transient magnetic anisotropy: 
 $e$--$h$ excitations
(non--thermal and coherent carrier contribution, left part) 
and 
Fermi sea holes 
(thermal  contribution,  
right part).
For $\hbar \omega_p$$\sim$3.1eV,   the 
holes are excited in 
high--${\bf k}$, 
non--parabolic,  
 HH or LH exchange--split valence 
band states.
(b): The thermal hole Fermi sea 
free energy gives four in--plane magnetic memory 
states $X^+$, $Y^+$, $X^-$, and $Y^-$, 
slightly tilted from the corresponding 
crystallographic axes. 
 } 
\label{Fig1}
\end{center} 
\end{figure*}

As is known in both semiconductors 
\cite{Chemla,rossi,mukamel,cundiff,chovan} and metals, 
\cite{petek,fann,taylor}
depending on the timescale, 
a distinction must be made
 between  $e$--$h$ quantum excitations, 
non--thermal $e$ and $h$ populations, and 
Fermi--Dirac populations (see the schematic 
in Fig. \ref{Fig1}(a)).
Initially, only coherent $e$--$h$  pairs 
are photoexcited 
(left part of Fig. \ref{Fig1}(a)).
  At a second stage,  scattering events
lead to the decay  of these 
 quantum excitations within a time--interval $T_2$. 
When $T_2$$<$100fs, 
such dephasing occurs during the laser pulse
and the  treatment of 
$e$--$h$ 
coherence is necessary only for  describing the 
 nonlinear fs photoexcitation
processes. The contribution of 
$e$ and $h$  populations  of the photoexcited states  
 is important 
when their relaxation/thermalization  time $T_1$ is 
not too short  compared to the $\sim$100fs 
timescales of interest. \cite{fann}
Such 
non--thermal  populations  
redistribute  between the band states 
as they relax
into hot  Fermi--Dirac distributions within $T_1$ 
(Fig. \ref{Fig1}(a)). \cite{fann} 
This relaxation 
 occurs 
after  picosecond (ps) times
(via emission of multiple phonons)
or faster (10's to 100's of fs via  
Coulomb interactions).

While the quantum  kinetics  
of charge photoexcitations 
is well--studied, 
\cite{rossi,fann} 
fs non--adiabatic magnetic correlation  
 is not well--understood.  
\cite{li-2013,Li-2,Mentink,kapetanakis-2009}
Collective spin dynamics 
is triggered  when  coupled magnetic order parameter
components  
are ``suddenly'' brought out of equilibrium
during coherent photoexcitation. 
Photoinduced  coherent, non--thermal,
and hot--Fermi--Dirac mobile carrier spins  
interact 
with the collective 
local spin 
during ultra--short timescales.  
Their  relative 
contributions
depend on laser intensity and  frequency, 
relaxation parameters, 
material  properties,
 and probed
timescales. 
 Pulse--shaping 
\cite{pulseshaping} 
and 
sequences  of fs laser--pulses
 analogous to multidimensional NMR
 spectroscopy \cite{NMR,cundiff}
offer additional possibilities for 
clarifying  and controlling 
such transient magnetic response.
Here we 
show
that coherent optical control of 
non--equilibrium mobile  carrier spin  induced by 
non--thermal population imbalance 
can suppress or start 
magnetization ringing or switching rotation 
at any time, by exerting  {\em fs spin--orbit torque}
 sequences 
in the appropriate directions. 
In this 
non--adiabatic way, 
we can
control magnetic 
states
without relying on
magnetic field pulses, circularly--polarized light,
\cite{kiriluk,chovan,nemec-2012} 
demagnetization, \cite{wang-rev,theory,beau,demag}
precession phase, \cite{Hashimoto-apl}
  or quasi--thermal processes.
\cite{aesch,Radu,jong,bigot-chemphys,qi}

The fs
 photoexcitation of (Ga,Mn)As 
has 
revealed different transient 
magneto--optical responses, such as 
ultrafast increase (decrease) of magnetization amplitude under
 weak (strong) excitation 
\cite{wang-rev,wang-2007,theory} 
and magnetization re--orientation 
due to spin--torque 
\cite{nemec-2012,chovan} 
 and spin--orbit torque.
\cite{wang-2009,nemec-2013,kapetanakis-2009}
There is mounting 
evidence that 
 non--thermal magnetic processes 
play an 
important role 
in the  fs magnetization time evolution.
\cite{kapetanakis-2009,chovan,wang-2009,nemec-2013} 
Such (III,Mn)V  heterostructures 
are advantageous for   optical  control 
of magnetic order 
due to their  well--characterized optical 
and electronic properties and manipulable 
carrier--induced  ferromagnetism.
Useful for demonstrating 
our theoretical predictions 
is that 
these systems have  four
different in-plane magnetic states
($X^+$, $Y^+$, $X^-$, and 
$Y^-$)
due to bi--axial magnetic anisotropy 
between the [100] and [010]  crystallographic axes
 (Fig. \ref{Fig1}(b) and Appendix \ref{nonadia}).
While  in conventional ferromagnets
switching involves 
spin--flipping  between two magnetic states
(spin--up/spin--down),   
the existence of four
 magnetic states 
allows for more complex multi--state 
switching pathways. Their experimental observation
can validate more elaborate  magnetization  
coherent control schemes, such as the one 
proposed here.
Four--state magnetic memories 
are also useful for 
ultra--high--density magnetic recording.
Two  equivalent 
 easy axes 
double the recording density, by  recording 
two bits of 
information on the same spot.
\cite{asta}
To take advantage of multi--state magnetic
memories for ultrafast spintronics applications, 
we must be able to 
selectively access all 
magnetic states 
in any      desired     sequence. 
There is no  generally accepted scheme for this. 
However,  optical spin manipulation has already reached a 
high level of sophistication 
\cite{wang-rev,kiriluk,Hashimoto-apl,asta,qi,dmwang,wang-2009,kapetanakis-2009,chovan,kapet-corr,axt,reiter} and 
control of magnetization   on a
100ps timescale
has been demonstrated in various systems, by   
using magnetic field or
laser--generated magnetic pulses 
\cite{gerrits,schumacher,kaka} 
or photoinduced effects. 
\cite{carpene,jong} 
Two outstanding challenges  have to be better addressed: 
(i) how to initiate and stop 
controlled deterministic switchings
during fs time intervals, 
(ii) how to suppress the magnetic ringing associated with 
switchings, which limits the prospects
for high--speed applications. \cite{stohr} 
From a more general perspective, 
the dynamical disentanglement of  
 degrees of freedom that are 
strong--coupled  
in equilibrium, e.g. the mobile and local 
collective spin sub--systems studied here, 
may lead to 
 a better understanding 
of correlated systems. \cite{Porer,wang-nc,Li-2,li-2013} 
The advantage  of using
spin--charge quantum kinetics 
to overcome the  limitations of 
incoherent processes
for meeting the above challenges 
is now beginning to be recognized.
\cite{li-2013,Li-2,axt,kapetanakis-2009,kapetanakis-2011,chovan,reiter,wu}

This work contributes to the 
debate  of how fs
coherent  photoexcitation 
could drive and control  ultrafast 
switchings
 \cite{li-2013,bigot1} and magnetic ringing. \cite{stohr}
 We consider the very early non--thermal and coherent 
temporal regimes 
and  focus mostly on 
 magnetization changes 
 {\em during the fs laser pulse}. 
We 
show that, by choosing  appropriate sequences 
of time--delayed 
laser pulses,  
we can 
control
 the   direction, magnitude, and time--profile
of the short--lived non--thermal photocarrier  spin. 
The latter drives the 
magnetization away from equilibrium 
by exerting  
fs
spin--orbit  torque on the 
collective local spin.  
By coherent manipulation of the  $e$--$h$ photoexcitations, 
we 
photogenerate  a  controlled population 
imbalance between 
  spin--orbit--coupled/exchange--split 
bands.
Such photoexcited 
band carrier  population and spin imbalance  
is not restricted by the chemical potential or temperature 
and 
leads 
 to a fully controllable 
 ``sudden''   magnetization 
canting  in selected  directions
 at desirable times. 
Based on  direct control of the above non--thermal processes 
by the optical field, 
we
propose possible protocols that 
drive 
complex 360$^{o}$ magnetization  pathways  involving sequential  
90$^{o}$ deterministic switchings 
between  {\em four different} 
magnetic memory states.
Such spin control, as well as
suppression of both magnetic ringing 
and  switching rotations, 
 are  possible without circularly--polarized light
due to  relativistic spin--orbit coupling 
of the photocarriers.
For linearly--polarized fs optical  pulses,
 we show that the photoexcited carrier 
spin direction and amplitude 
is determined by 
the competition between  
spin--orbit coupling with characteristic energy 
$\Delta_{so}$$\sim$340meV given   by the $\Gamma$--point
energy splitting
 of the GaAs spin--orbit--split valence band,
and 
the ${\bf S} \cdot {\bf s}$ magnetic 
exchange coupling, with characteristic 
energy $\Delta_{pd}$=$\beta c S$$\sim$100meV
in Ga(Mn)As,  \cite{jung} where 
$S$ and $c$ denote the Mn spin amplitude and concentration
respectively 
and  $\beta$ is the magnetic exchange constant.
The time--reversal  symmetry breaking 
can be characterized by the  energy ratio 
$\Delta_{pd}$/$\Delta_{so}$($\sim$1/3 
in (Ga,Mn)As).
It leads to fs 
photoexcitation  of 
 short--lived mobile spin--pulses
(${\bf s}$), whose  
direction 
is  controlled 
by   selectively populating 
the continua of 
exchange--split 
heavy--hole (HH) or light--hole (LH) 
spin--orbit--coupled  band states  
with different spin superpositions.
We model the fs 
nonlinear photoexcitation  processes, driven 
by sequences of time--delayed laser--pulse--trains,  
with density matrix equations--of--motion  \cite{rossi}
describing  carrier populations coupled non--perturbatively to 
inter--band coherences and local spins.
Our time--domain calculations describe
 a  non--equilibrium 
magnetic anisotropy 
during the laser pulse, which we 
  estimate 
by  treating  strong 
band non--parabolicity and spin--orbit couplings  
using the tight--binding bandstructure  
of GaAs with 
 mean--field 
magnetic exchange interaction. \cite{jung,vogl}
We relate the calculated 
coherent photoexcitation 
of fs spin--orbit torque  
to existing experiments 
 and make predictions for 
new ones to
 observe  switchings
by using pulse--shaping.\cite{pulseshaping}

The paper is organized as follows.
In Section \ref{nonthermal} we discuss the  
symmetry--breaking
processes
leading  
to  photoexcitation  
of a 100fs mobile 
carrier spin--pulse  
 with direction and magnitude
that depend on the ratio 
$\Delta_{pd}/\Delta_{so}$.
In Section \ref{Init} we compare theory and experiment 
to demonstrate coherent control of 
 fs spin--orbit torque 
direction and magnitude 
 by tuning 
 populations of four exchange--split 
HH and LH valence bands excited by a 
laser pulse.
We   show that 
the canting direction of a transverse,  out--of--plane, 
 fs magnetization 
component
displays a  magnetic hysteresis
absent without pump, which is distinguished 
from 
longitudinal amplitude and nonlinear optical effects
by sweeping a perpendicular magnetic field.
In Section  \ref{one-train}  we show that 
we can initiate controlled  switching rotations to any one of the 
available magnetic states
 by shaping a laser--pulse train.
In Section \ref{Protocols} we propose two
protocols for controlling 
four sequential 90$^{o}$ switchings in clockwise or
counter--clockwise directions.
In Section \ref{two-train} 
 we use  two time--delayed laser--pulse--trains 
to  suppress or enhance the nonlinear switching rotation
at any intermediate state and  to suppress 
magnetic ringing at any time,  long or short.
Rather than relying  on the 
magnetization precession phase, 
we achieve this 
coherent control by switching the {\em directions}
of  fs spin--orbit torques.
We end with conclusions and a broader outlook.
In two Appendices we 
present the 
density matrix equations
describing nonlinear coherent excitation of 
fs spin--orbit torque,  separate
 non--adiabatic/non--thermal 
 and adiabatic/thermal 
transient magnetic anisotropy,
and treat the  non--parabolic and 
anisotropic spin--orbit--coupled 
 band continua.

\section{Femtosecond spin photoexcitation} 

\label{nonthermal}

In this section  we 
discuss the general 
processes leading to 
photoexcitation
of carrier spin with direction determined by 
non--perturbative 
 symmetry--breaking interactions.
In the systems of interest,
 the magnetic effects arise from  antiferromagnetic  
 interactions between 
localized and mobile (delocalized) carrier spins. \cite{jung} 
In contrast to   magnetic
insulators studied before, \cite{kiriluk} 
 the 
 localized electrons 
do not contribute to the 
fs magnetic anisotropy. They determine 
the magnetization (collective local spin) 
\begin{equation} 
{\bf S} = \frac{1}{c V} \sum_{i} \langle {\bf \hat{S}}_i \rangle,
\label{S-def} 
\end{equation} 
where $V$ is the volume
and ${\bf S}_i$ are the local magnetic moments
at 
positions $i$, with  
concentration $c$.   
For example, in (III,Mn)V 
magnetic semiconductors, 
 the local magnetic  moments  
are
pure $S$=5/2 Mn
spins with zero angular momentum, $L$=0, and 
no spin--orbit interaction. 
The fs magnetic anisotropy comes from 
 band electrons that are 
subject to spin--orbit interactions and,
unlike for the    localized 
electrons,
couple directly to light. 
The spin--exchange  coupling of such photoexcited 
itinerant carriers
with the local spins
results in photoinduced magnetization dynamics.
The widely--used  mean--field treatment of the 
magnetic exchange interaction (Zener model) 
captures the 
symmetry--breaking 
of interest here. \cite{jung}
We thus 
consider the dynamics of a single--domain macrospin ${\bf S}$(t)
and neglect  
spatial fluctuations. \cite{kapet-corr,axt} 
This approximation
describes metallic--like
(III,Mn)V and other magnetic semiconductors.   \cite{jung}

Our main goal here is to control the 
non--equilibrium spin of  band carriers
in order to manipulate the magnetization motion.
These non--equilibrium  carriers
dominate the laser--induced 
magnetization changes during fs timescales. 
While spin--lattice coupling 
affects the easy axis, 
lattice heating and relaxation 
occurs on  ps timescales, 
following energy transfer from  electronic system 
of interest here.  
\cite{nemec-2013,qi}
The laser
excites $e$--$h$ pairs   between the
different  exchange--split valence and 
conduction bands
(Fig. \ref{Fig1}(a)).  
Magnetic exchange 
mainly involves the valence hole  collective 
 spin  
${\bf s}_{h}$.
Denoting by 
${\bf s}_{{\bf k} n}^{h}$ 
the  contribution 
of valence band $n$
with 
 given  ${\bf k}$,
\begin{equation} 
{\bf s}_{h}(t)=
\frac{1}{V}
\sum_{{\bf k}} \sum_{n} 
{\bf s}_{{\bf k} n}^{h}(t).
\label{net-spin} 
\end{equation} 
We want to
 control 
${\bf s}_{{\bf k} n}^{h}(t)$ 
between  different  bands.
We describe this spin 
by extending 
the discrete--${\bf k}$ calculation of 
fs spin--orbit--torque
in
Ref.[\onlinecite{kapetanakis-2009}]
to include the anisotropic continua of non--parabolic 
(Ga,Mn)As bands and to consider 
sequences of laser--pulse--trains. 
We can thus estimate  the 
photocarrier 
density and net spin
as function of 
 laser--pulse frequency, intensity, and time delays
for comparisons to experiments.
The  
mechanism of  Ref.[\onlinecite{kapetanakis-2009}]  
is analogous to the 
current--induced 
spin--orbit torque \cite{zhang} 
observed  in (Ga,Mn)As \cite{furdyna} 
and other spin--orbit--coupled ferromagnets. 
Unlike in our
earlier 
work \cite{chovan}  on fs
 spin--transfer torque analogous to the one induced by
spin--polarized 
currents in spintronics 
applications,
\cite{berger,brataas} 
which is observed experimentally by using circularly--polarized 
light, \cite{nemec-2012}
here  
spin is not conserved due to the  spin--orbit coupling.
As a result,  angular momentum transfer 
from the photons is not necessary
due to the symmetry--breaking  provided 
by the competition of spin--orbit and magnetic exchange 
couplings. 

To initiate  ultrafast dynamics, 
we  create a short--lived spin imbalance 
by 
optically controlling 
the individual 
contributions  ${\bf s}_{{\bf k} n}^{h}$(t)
 of different electronic 
bands  and Brillouin zone directions.
To describe this, we
express the mobile carrier spin 
in terms of the electronic 
density matrix
$\langle \hat{h}^{\dag}_{-{\bf k}n}\hat{h}_{-{\bf k}n'}\rangle$,
defined 
in terms of  an adiabatic 
basis
of band  eigenstates
 created by the operators $\hat{h}^\dag_{-{\bf k}n}$:
\begin{equation}  
\ {\bf s}_{{\bf k} n}^{h}= 
\hat{{\bf s}}^h_{{\bf k}nn}
\, \langle \hat{h}^{\dag}_{-{\bf k}n}\hat{h}_{-{\bf k}n}\rangle
+ 
\sum_{ n' \ne n }
\hat{{\bf s}}^h_{{\bf k}nn'}
\, \langle \hat{h}^{\dag}_{-{\bf k}n}\hat{h}_{-{\bf k}n'}\rangle,
\label{h-spin} 
\end{equation} 
where 
$\hat{{\bf s}}^h_{{\bf k} n^{\prime} n}$ 
are the spin matrix elements. These 
describe the spin direction
of carriers populating 
band states $(n,{\bf k})$. 
Spin direction 
changes for different
${\bf k}$--direction  and 
band.
It
is determined by
spin--mixing due to 
non--perturbative interplay of
 spin--orbit and magnetic exchange couplings, which depends 
on $\Delta _{pd}/\Delta_{so}$.  
The first term  on the right hand side (rhs) of 
Eq.(\ref{h-spin}) 
describes the population contribution 
(coherent, non--thermal, or quasi--thermal). 
 The second term descibes the contribution due to 
coherent coupling  of different bands
(inter--valence--band coherence). Such Raman coherence  
arises  
because 
spin is not conserved, 
$\hat{{\bf s}}^h_{{\bf k}nn'} \ne$0, 
and vanishes in equilibrium.
We choose as basis  $\hat{h}^\dag_{-{\bf k}n}$
 the eigenstates ($n$,${\bf k}$)  
of the 
 adiabatic Hamiltonian 
(Appendix \ref{nonadia})
\begin{equation} 
H_{b}({\bf S}) 
=H_{0}+H_{so}+H_{pd}({\bf S}_0).
\label{Hb} 
\end{equation} 
  $H_{0}$+$H_{so}$ 
describes the bandstructure 
of the parent 
material (undoped GaAs here), due to 
the 
periodic lattice potential ($H_0$) and the 
spin--orbit coupling
($H_{so}$).\cite{vogl} 
The symmetry--breaking is induced  by 
the  magnetic 
exchange interaction 
$H_{pd}({\bf S}_0)$, Eq.(\ref{DH}). 
\cite{jung}  
Here,  ${\bf S}_0$ denotes the 
slowly--varying contribution to the local macrospin,  
 which  switches 
or oscillates during ps timescales
(adiabatic approximation). 
The valence hole and conduction electron basis  states, 
$\hat{h}^\dag_{-{\bf k}n}$ and 
$\hat{e}^\dag_{{\bf k}m}$ respectively, 
were obtained  by diagonalizing $H_b$(${\bf S}_0$)
using the 
tight--binding 
approximation of Ref.[\onlinecite{vogl}]
(Appendix \ref{nonadia}). 

In (III,Mn)V semiconductors, 
 a thermal hole Fermi sea bath, 
characterized by the Fermi--Dirac 
distribution $f_{n {\bf k}}$, 
 is already present 
 in the  ground state (Fig. \ref{Fig1}(a)).
 \cite{jung}  
Similar to  ultrafast studies of the 
electron gas in metals \cite{fann} and semiconductors, 
\cite{Perakis,Karadim}
we distinguish this 
quasi--equilibrium  contribution to 
Eq.(\ref{h-spin})
from the  non--Fermi--Dirac  
femtosecond contribution  
 (Appendix \ref{nonadia}):
\begin{equation} 
\langle \hat{h}^{\dag}_{{\bf k}n}
\hat{h}_{{\bf k}n'}\rangle
= \delta_{n n^\prime} 
f_{n {\bf k}} + 
\Delta \langle \hat{h}^{\dag}_{{\bf k}n}
\hat{h}_{{\bf k}n'}\rangle.
\label{holepop} 
\end{equation} 
At quasi--equilibrium,
only the Fermi--Dirac populations
contribute. These 
are characterized by a 
 temperature 
and  chemical potential  
and give  the adiabatic 
 field  
\cite{kiriluk,nemec-2013,jung}
\begin{equation} 
\gamma {\bf H}_{ FS}[{\bf S}]=-
\frac{\partial E_h({\bf S})}{\partial {\bf S}}, 
\label{HFS}
\end{equation} 
where $\gamma$ is the gyromagnetic ratio 
and 
\begin{equation}
E_h({\bf S})= 
\sum_{{\bf k} n} \, \varepsilon^v_{n {\bf k}} \,
 f_{n {\bf k}}
\label{Eh-eq} 
\end{equation} 
is the total (free) energy 
of the relaxed Fermi--Dirac carriers. 
$\varepsilon^v_{n {\bf k}}({\bf S})$ 
are the (valence band) eigenvalues of the 
adiabatic Hamiltonian $H_b$ for frozen local spin ${\bf S}$. 
The above adiabatic free energy of the mobile hole Fermi sea 
defines  the  magnetic memory states
of  Fig. \ref{Fig1}(b)
(Appendix \ref{nonadia}).    
Since the changes of $E_h$
with ${\bf S}$
are notoriously  small 
for numerical calculations 
of the quasi--equilibrium magnetic anisotropy, 
 \cite{anis,nemec-2013}  
while the low--energy states 
of (III,Mn)V systems 
are complicated by sample--dependent disorder, 
impurity bands, defect states, 
and strain, 
\cite{dmwang,theory,welp,jung} 
 we
approximate 
$E_h({\bf S})$  by 
using  the symmetry--based 
Eq.(\ref{eq:Eh})
with 
parameters extracted  from
experiment (Appendix \ref{nonadia}).
\cite{dmwang,welp,jung}
In this way,
we 
introduce the realistic  
four--state 
magnetic memory 
of the (III,Mn)V materials. 
For the low 10--100$\mu J$/cm$^{2}$ 
pump fluences considered here,  
we neglect 
any laser--induced changes in the 
 Fermi--Dirac distribution temperature and chemical potential, 
which  add to the predicted effects 
on the timescale of energy and 
population relaxation. 
Previous calculations 
assuming 
Fermi--Dirac distributions \cite{nemec-2013,theory}
gave order--of--magnitude 
smaller magnetization dynamics 
compared to experiment 
and concluded that the 
non--equilibrium hole distribution 
is very broad. \cite{theory} 
Here we study the possible role 
of 
the short--lived ($\sim$10--100fs) 
non--Fermi--Dirac  
populations that exist prior to full electronic 
thermalization.
We calculate such fs anisotropy 
 in the time domain, 
by solving
equations of motion for 
$\Delta \langle \hat{h}^{\dag}_{{\bf k}n}
\hat{h}_{{\bf k}n'}\rangle$
using the time--dependent 
Hamiltonian  (Appendix \ref{nonadia}) 
\begin{equation} 
H(t)=H_{b}({\bf S}_0) 
+\Delta H_{exch}(t)+H_{L}(t).
\label{H}
\end{equation} 
While $H_b$(${\bf S}_0$) changes 
during 10's of ps,
the other two  contributions to 
Eq.(\ref{H}) 
are non--adiabatic and 
vary  during 
fs timescales.
$H_{L}(t)$, Eq. (\ref{HL}), 
 describes the 
dipole coupling of the fs laser $E$--field, 
 \cite{rossi} while
\begin{equation} 
\Delta H_{exch}(t)=
\frac{1}{V} \sum_{{\bf k}} 
\beta_{{\bf k}} c \, \Delta {\bf S}(t)
\cdot \hat{s}^h_{{\bf k}},
\label{DHexch}
\end{equation} 
where $\hat{s}^h_{{\bf k}}$ 
is the hole spin operator 
and 
\begin{equation}
\Delta {\bf S}(t) = {\bf S}(t) - {\bf S}_0.
\label{DS}
\end{equation}
describes  ``sudden''  
 changes in magnetization
during fs photoexcitation. 
We assume 
exchange constant $\beta_{{\bf k}}$$\approx$$\beta$ 
for simplicity within 
 the relevant range of ${\bf k}$.
The effects of  non--adiabatic
$\Delta {\bf S}$(t)
on the time--dependent  band states 
are described via the density 
matrix equations of motion.

We describe the non--Fermi--Dirac electronic contribution 
$\Delta \langle \hat{h}^{\dag}_{{\bf k}n}
\hat{h}_{{\bf k}n'}\rangle$, Eq.(\ref{holepop}),  
similar to the well--established 
Semiconductor Bloch Equation \cite{koch,rossi}
or
 local--field  
\cite{mukamel,mukamel1} 
Hartree--Fock
treatments of ultrafast 
nonlinear optical  response.  
In particular, we  solve coupled 
equations 
of motion
for the electronic populations and inter--band  coherences 
$\langle
 \hat{h}^\dag_{{\bf k} m} 
\hat{h}_{{\bf k} n} \rangle$,  
 $\langle \hat{e}^\dag_{{\bf k} m} 
\hat{e}_{{\bf k} n} \rangle$,  
and $\langle
 \hat{e}_{{\bf k} m} 
\hat{h}_{-{\bf k} n} \rangle$ 
that are   non--perturbatively 
coupled
to the time--dependent local spin ${\bf S}$(t). 
This coupling  modifies the electronic dynamics, which, in turn, 
modifies the motion of 
${\bf S}$(t)
(Appendix \ref{nonadia}).
We consider linearly--polarized optical 
pulses 
with zero angular momentum.
We do not include  the carrier-carrier, 
carrier-phonon, and carrier--impurity interactions
in the Hamiltonian, 
but treat the photocarrier relaxation phenomenologically, 
by introducing $e$--$h$ lifetimes 
(dephasing time $T_2$) 
and non--thermal population
relaxation times 
$T_1$.
Our calculation thus 
describes the ``initial condition'' that brings the system 
out of equilibrium and initiates relaxation.  
\cite{wu,theory} 
The latter 
redistributes the non--thermal  carriers 
among band states with different spins 
and momentum directions ${\bf k}$. 
Here we introduce a relaxation time $T_1$ 
for the populations 
$\langle \hat{h}^{\dag}_{-{\bf k}n}\hat{h}_{-{\bf k}n}\rangle$
that determine the hole spin in Eq.(\ref{h-spin}),  
which thus reflects the hole spin relaxation time. 
The latter  
was calculated in Ref.[\onlinecite{wu}]
to be several 10's of fs and was measured 
in (Ga,Mn)As 
to be in the range of $\sim$200fs. \cite{spinrelax}  
Momentum scattering and carrier relaxation is also expected to 
give 
$T_2$ of  few 10's of fs. \cite{jung,wu} 
Below we estimate 
the dependence of the predicted non--thermal effects 
on  $T_1$ and $T_2$. 

The present calculations
describe the
spin photogeneration 
that initiates the dynamics.
Our main focus 
is on  the average
hole spin  $\Delta {\bf s}_h$(t) coming 
from 
 $e$--$h$ 
photocarriers, whose 
non--Fermi--Dirac
population of band 
continuum states 
depends on the 
laser frequency $\omega_p$.
The  results presented here were obtained for 
$\hbar \omega_p$$\approx$3.1eV. \cite{wang-2009,burch} 
For such high frequencies, the 
disorder--induced  impurity/defect states 
\cite{theory}   
do not contribute significantly  and 
the 
photoexcited  carriers 
are initally 
well--separated in energy 
from the Fermi sea holes
(see Fig. \ref{Fig1}(a)). 
We
 mainly excite band states along the 
eight \{111\} symmetry lines  of the Brillouin zone, 
at high ${\bf k}$,  
where the conduction and valence bands are strongly 
non--parabolic and 
almost parallel to each other. \cite{burch} 
As a result, 
a large number of 
inter--band 
optical 
 transitions 
are excited and 
 a broad range of band momenta  ${\bf k}$ 
is populated 
(see Fig. \ref{Fig1}(a)). 
The  effects of such highly anisotropic 
band continua on the 
photoexcited hole spin 
 are accounted for
as described in Appendix \ref{cont}.
Due to the symmetry--breaking introduced by 
${\bf S}$(t), 
the eight 
 \{111\} directions not equivalent, unlike in 
GaAs, 
which leads to anisotropy.  
Assuming  smoothly varying exchange constant $\beta_{{\bf k}}$, 
the hole spin matrix element  
$\hat{{\bf s}}^h_{{\bf k}nn'}$ is 
fairly constant over a wide range of ${\bf k}$,
but differs between the eight photoexcited 
\{111\}  directions. 
In this case, the contribution to the 
average spin $\Delta {\bf s}_h$(t) 
from each band $n$ and each direction of ${\bf k}$
 is  approximately proportional to 
the photoexcited densities
$
\frac{1}{V} \sum_{k} 
\Delta \langle \hat{h}^{\dag}_{-{\bf k}n}\hat{h}_{-{\bf k}n}\rangle
$. 
Our calculation estimates 
these anisotropic total densities 
prior to full inter--band relaxation and 
large momentum scattering  
between different symmetry directions
leading to hole spin relaxation.
By tuning the laser frequency 
$\omega_p$, we create a short--lived  imbalance between 
different bands $n$, whose spin--admixture differs 
due to the different spin--orbit interactions. 
 Fermi--Dirac populations of relaxed non--equilibrium holes 
 add to the 
 $\Delta {\bf s}_h$(t) calculated here.
For 
 $\hbar \omega_p$$\approx$1.5eV,
\cite{nemec-2013} 
one  excites smaller ${\bf k}$ along all
\{100\}, \{010\}, 
\{001\}, \{110\}, \{101\}, \{011\}, and 
\{111\} symmetry 
 directions, 
 \cite{enders} 
as well as a distribution of 
impurity/defect states 
inside the semiconductor bandgap.\cite{theory,jung} 
Despite this difference, 
the qualitative conclusions of our work apply 
to all frequencies. The main requirement 
is to be able to coherently  induce 
a  non--equilibrium 
 population imbalance  
between different 
bands and momentum directions, 
which here is
facilitated by the symmetry breaking.

Important for  bringing the 
local and mobile spin sub--systems away from equilibrium 
is the  difference in their ultrafast  dynamics.  
Unlike for the band carriers, 
 there is no spin--orbit or optical field coupling
of the  local spins here.
In equilibrium, 
the 
local and mobile collective spins 
are correlated 
in the ferromagnetic state, 
so that ${\bf S} \times {\bf H}_{FS}$=0.
\cite{jung}
Within the mean--field approximation,
${\bf S}$(t)
is 
 driven 
 out of this  equilibrium configuration 
according to a  
Landau--Lifshitz--Gilbert 
equation.
\cite{kiriluk}
The latter describes a
magnetization re--orientation
driven by  both  quasi--equilibrium 
(${\bf H}_{ FS}$) and  
  non--thermal
($\Delta {\bf s}_h$)  mobile 
carrier 
spins, which modifies the electronic states:
\begin{equation}
\partial_{t}{\bf S}= 
-\gamma {\bf S}\times {\bf H}_{ FS}[{\bf S}(t)] 
-\beta{\bf S}\times
\Delta {\bf s}_h(t) 
+\frac{\alpha}{S} {\bf S}\times \partial_{t}{\bf S},
\label{eom-Mn}
\end{equation}
where 
$\alpha$ characterizes the slow
local spin precession damping.
\cite{qi}
Laser--induced 
 magnetization amplitude changes are not captured by this 
mean--field approximation and  
require  treatment of  spin--charge 
correlations  described in 
Refs.[\onlinecite{theory,axt,kapet-corr,Li-2}].
By including such correlations, 
longitudinal demagnetization effects
triggered by the non--equilibrium population imbalance 
give a magnetization amplitude 
$S$(t), which is distinguished 
from the transverse effects 
of main interest here.

The dynamics of the 
 mobile carrier spin
depends not only on the magnetic exchange 
interaction with the local spin but also 
on spin--orbit coupling, direct 
nonlinear coupling to the optical 
field,  
and  fast relaxation: 
\cite{chovan}
\begin{equation} 
 \partial_t {\bf s}^h_{{\bf k}}=
\beta c{\bf S}\times{\bf s}^h_{{\bf k}}
+ i
\langle [H_{so},{\bf s}^h_{{\bf k}}] \rangle
+Im  \, {\bf h}_{{\bf k}}(t) 
+ \left. \partial_t {\bf s}^h_{{\bf k}} \right|_{rel}.
\label{hole-spin} 
\end{equation} 
The above equation 
is  not useful here, 
as it does not distinguish between 
different bands in order to treat 
the spin--orbit coupling
$H_{so}$. 
Nevertheless, it 
demonstrates the 
four  processes  
that determine the non--thermal carrier  spin.
The first term describes spin--torque 
 due to  magnetic exchange.
The second term describes  {\em spin--orbit 
torque}, obtained  here 
by calculating the 
density matrix Eq.(\ref{holepop}).
The third term 
describes 
Raman--type 
coherent  nonlinear optical processes
that 
drive  photoexcitation 
 of band carrier spin
with symmetry--breaking:\cite{chovan} 
\begin{eqnarray} 
{\bf h}_{{\bf k}}(t) = 
 2  \sum_{m n} \, 
\langle \hat{h}_{-{\bf k}
n} \hat{e}_{{\bf k} m} \rangle
\, 
\sum_{m^{\prime}}
 {\bf d}_{{\bf k} m m^{\prime}}^*(t) 
\cdot {\bf s}^h_{{\bf k} m^{\prime} n},
\label{h-def}
\end{eqnarray}
where 
$ d_{{\bf k} m m^{\prime}}(t)$ 
=${\bf \mu}_{{\bf k} m m^\prime} \cdot {\bf E}$(t)  
are the Rabi energies of
 optical transitions between band states 
$(m{\bf k})$ and $(m^\prime{\bf k})$
and  ${\bf E}$(t) is the laser $E$--field.
Finally, the last term 
 describes the fast spin relaxation 
due to scattering processes. \cite{wu,chovan}

\begin{figure} [t]
\begin{center} 
\includegraphics[scale=0.3]{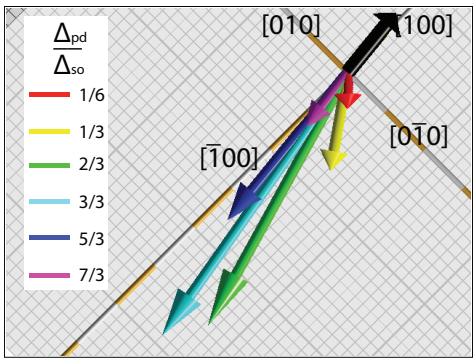} 
\caption{(Color online)
Maximum of 
anisotropy field 
 $\beta \Delta {\bf s}_h$(t) 
  photoexcited by a single 100fs linearly--polarized laser pulse,  
as function of  energy ratio $\Delta_{pd}/\Delta_{so}$
characterizing
time--reversal symmetry breaking. 
The direction of the ground--state magnetization is along the 
X$^{+}$ easy axis represented by the black arrow close to [100].
$\hbar \omega_p$=3.14eV, $E_0$=7$\times$10$^5$V/cm, 
$T_1$=100fs, $T_2$=50fs.
} 
\label{Fig2}
\end{center} 
\end{figure}

Spin--orbit and magnetic exchange couplings 
play an important role 
during fs photoexcitation: their non--perturbative 
interplay determines the 
direction and magnitude of the photoexcited carrier spin. 
The strong dependence of 
the 
maximum of the photoexcited hole fs spin--pulse 
 $\beta \Delta {\bf s}_h(t)$
on the
energy ratio $\Delta_{pd}/\Delta_{so}$
is  demonstrated by Fig. \ref{Fig2}.
This result was  
obtained by solving the 
coupled equations of motion 
discussed in Appendix \ref{nonadia}.
Fig. \ref{Fig2} demonstrates a strong dependence 
of photohole net spin direction and magnitude 
on 
$\Delta_{pd}/\Delta_{so}$.
In the ground state, the magnetization 
${\bf S}_0$
points along the 
$X^+$ easy axis close to [100]
(Fig. \ref{Fig2}). 
For $\Delta_{pd}$$\ll$$\Delta_{so}$, 
$\Delta {\bf s}_h$ is negligible without 
 circularly--polarized light.
The average spin 
vanishes  
since 
all symmetric directions in the Brillouin zone are 
excited 
equally.
With increasing $\Delta_{pd}$,  the 
magnetic  exchange interaction
 introduces  a preferred direction
along ${\bf S}$(t), which 
 breaks the time--reversal symmetry of GaAs
and results in a net 
 $\Delta {\bf s}_h$(t) 
while the laser pulse couples 
to the magnetic system.
With increasing $\Delta_{pd}/\Delta_{so}$, 
the magnitude of this  
$\Delta {\bf s}_h$ 
increases and its direction changes. 
For $\Delta_{pd}/\Delta_{so}$$\sim$1/3,
as in  (Ga,Mn)As, 
Fig. \ref{Fig2} shows  that
 the 
in--plane component of the 
fs 
anisotropy field 
$\beta \Delta {\bf s}_h$ points 
close to the $[\bar{1}\bar{1}0]$ direction
for $\hbar \omega_p$=3.14eV.
As discussed below, this result  is consistent 
with the experimental observations.  
The above $\Delta {\bf s}_h$(t)
 only lasts during the 100fs laser pulse and 
drives a ``sudden'' 
 magnetization canting 
$\Delta {\bf S}$(t) 
via fs spin--orbit torque. 
As $\Delta_{pd}$ approaches $\Delta_{so}$, 
 $\Delta {\bf s}_h$ is maximized
while it changes direction. It  
decreases again 
for $\Delta_{pd}$$\gg$$\Delta_{so}$.
In the next section we 
compare 
to  experiment.

\section{Exciting 
Spin Dynamics 
with a Single 
 pulse:
theory vs experiment} 
\label{Init} 

Ultrafast magneto--optical experiments
in (III,Mn)V semiconductors 
have revealed
 magnon oscillations with frequency $\Omega$$\sim$100ps$^{-1}$, 
which can be 
 suppressed
(enhanced) with a control laser pulse delayed 
by $\tau$  if 
$\Omega  \tau$=$\pi$ ($\Omega  \tau$=2$\pi$). 
\cite{Hashimoto-apl} 
In this paper 
we propose a different optical 
coherent control scheme, 
based on 
controlling the {\em direction, duration, and magnitude} 
of fs spin-orbit torque seuqences  
photoexcited  at any time $\tau$. 
We are not aware of any experiment so far
showing laser--induced 
 360$^{o}$  switchings between multiple magnetic 
states and suppression of
switching rotation 
at an arbitrary magnetic 
state.
In subsequent sections 
we explore 
 how to observe 
our theoretical prediction experimentally by using 
pulse--shaping. 
In this section,  we validate our original 
prediction of fs spin--orbit torque 
in  (III,Mn)V materials,  \cite{kapetanakis-2009} 
by comparing numerical results obtained 
for anisotropic, non--parabolic band continua 
with existing experiments
showing 
fs magnetic hysteresis
excited by a single 100fs laser 
pulse in (Ga,Mn)As.
It is certainly valuable to establish 
the connection of our theory  with this 
experiment  
 before  making  numerical
 predictions of complex protocols based on trains of 
laser pulses with various timing sequences and colors. 
 As discussed
 in subsequent sections, 
laser--pulse--trains 
provide a  more controlled way 
to enable  multiple switchings  
by manipulating 
fs spin--orbit torques. 
In this way we 
maximize    ``transverse''
hole spin excitations
while keeping the pump fluence as low as possible 
(10--100 $\mu$J/cm$^{2}$) 
to reduce the 
``longitudinal'' fs demagnetization. 
The direct theory--experiment 
comparison in this section, 
as well as the indirect connection to other experiments 
\cite{nemec-2013} 
discussed later,  
  makes the case 
that 
optical control of 
a  short--lived coherent 
population imbalance between 
exchange--split, spin--orbit--coupled anisotropic bands
can  generate fs spin--orbit torque 
with  controllable direction, temporal profile, and magnitude. 
 
\begin{figure}[t]
\begin{center}
\includegraphics[scale=0.36]{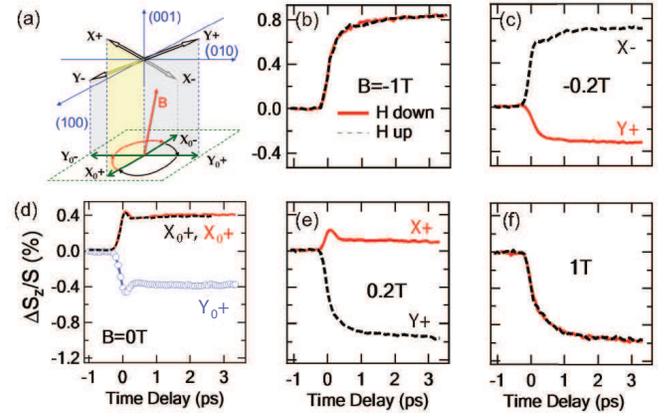}
\caption{(Color online) Magneto--optical pump--probe experimental 
measurements 
showing a step--like, laser--induced, 
fs magnetization canting  $\Delta S_z$(t)
that
displays a
fs  magnetic hysteresis and 
 switches direction  when switching 
magnetic  state. 
(a) An external perpendicular $B$--field, applied 
at small angle  $\sim$5$^{o}$ from the 
[001] axis, 
tilts the  $B$=0 in--plane 
easy axes ($X_0^{\pm}$ and $Y_0^{\pm}$)
out of the plane
(Appendix \ref{nonadia}).
(b)--(f): 
The ``sudden'' out--of--plane magnetization 
 $\Delta S_z/S$, 
induced by a  100fs laser pulse with  fluence 
$\sim$7$\mu J/cm^{2}$,  
switches direction 
when sweeping the above  B--field 
between 
$B$=-1T (b) and $B$=1T (f).
The two sweeping 
directions correspond to 
increasing (up, 
full curves) and
decreasing 
(down, dashed curves)
B--field. 
For each of the measured 
 $B$=1T, 0.2 T, 0 T, -0.2T, -1T,
the fs temporal profiles of 
 $\Delta S_z/S$  depend on the equilibrium magnetic state, which is
switched when sweeping the $B$--field.
}
\label{Fig3}
\end{center}
\end{figure}

Fig.\ref{Fig3}
shows the detailed fs temporal profile of the pump--probe 
 magneto--optical 
signal 
of our (Ga,Mn)As sample 
as function of perpendicular magnetic field.
Unlike previous  experiments that measured 
magnetization  dynamics 
on the ps timescale, 
Fig.\ref{Fig3} 
shows  the  initial $\sim$100fs 
 temporal 
regime. 
As we demonstrate below, this regime 
reveals  a sizable 
carrier--spin pulse with $\sim$100fs duration, 
which generates  a transverse 
magnetization component 
that  cannot arise from longitudinal 
nonlinear optical or demagnetization effects.  
The details of the  experimental methods and 
(Ga,Mn)As sample 
may be found in Ref.[\onlinecite{wang-2009}]. 
 The 100fs pump pulse, 
with central frequency 
$\hbar \omega_p$=3.1eV, 
 is well--separated in energy
from the 
probe  tuned 
at 1.55eV, which  minimizes pump--probe 
interference.
The pump 
optical   field, with  amplitude $E_0$$\sim$2$\times$10$^{5}$V/cm
and 
fluence 
$\sim$7$\mu$J/cm$^{2}$, 
excites a total 
photohole density  
of $n$$\sim$6$\times$10$^{18}$/cm$^3$. This is  a small 
 perturbation  
of the 
3$\times$10$^{20}$/cm$^3$
ground state hole density in our (Ga,Mn)As sample.
$\Delta S_z/S\propto\Delta \theta_K/\theta_K$
is probed  via the pump--induced Kerr rotation angle  
$\Delta \theta_K$
along the [001] direction.
 \cite{wang-rev} 
The 
magnetic origin of the measured signal 
is implied by two experimental observations. 
First, 
$\Delta \theta_K$ coincides  
with the  pump--induced 
ellipticity 
even during $\sim$100fs.
 \cite{wang-2009}
Second,  we observe a systematic 
$B$--field dependence  and  sign switching of $\Delta \theta_K$,
 absent in $\theta_K$ without  pump, which 
 directly correlates with the 
magnetic switchings observed in the 
static transverse 
Hall magnetoresistivity.
 \cite{wang-2009}
 We   measure a transverse 
fs 
magnetization component $\Delta S_z$(t),  
 perpendicular to the ground state magnetization,  
which switches direction  
with initial magnetic state.  
This transverse component 
is  suppressed by increasing the magnetic field
perpendicular to the sample plane and easy axes.  
The observed  ``sudden'' step--like temporal profile
indicates that spin re--orientation 
 completes during the laser pulse and thus is 
driven by $e$--$h$ photoexcitation.
Such fs time--dependence is clearly distinguished    
 from subsequent 
 ``slow'' 
magnon oscillations during $\sim$100ps times.
\cite{spinrelax}

To elucidate the magnetic origin of the measured 
$\Delta \theta_K/\theta_K$ 
 with $\sim$100fs duration, 
we apply a transverse  external magnetic field $B$ 
as in Fig.\ref{Fig3}(a). 
For $B$=0, the magnetic states
$X_{0}^{\pm}$ and $Y_{0}^{\pm}$  lie inside the 
plane (Fig. \ref{Fig1}(b)).
For $B$$\ne$0, Eq.(\ref{anis-1}) gives 
an  out--of--plane canting of  
$X^{\pm}$ and $Y^{\pm}$
(Fig. \ref{Fig3}(a)).   
 In the absence of pump,
the measured smooth change of static Kerr rotation angle 
$\theta_K$ as function of $B$--field, 
discussed in Refs.[\onlinecite{wang-2009,spinrelax}], 
reflects this
canting. 
 There is 
no magnetic hysteresis  of $\theta_K(B)$   
without the pump. 
This is unlike the behavior 
demonstrated by 
Fig. \ref{Fig3}, which  shows the changes 
in the fs temporal profile of    
$\Delta \theta_K/\theta_K$
induced 
by sweeping the magnetic field 
between $B$=$-1T$ and $B$=$1T$. 
By comparing the fs 
response between increasing (``up'')
and decreasing (``down'')
$B$--field, \cite{wang-2009}  
we observe in 
Figs. \ref{Fig3}(b) through (f) a
 magnetic 
hysteresis of the laser--induced
fs  change 
$\Delta \theta_K/\theta_K$. 
This magnetic hysteresis and sign switchings
coincide with the magnetic hysteresis 
measured in the static 
Hall magnetoresistivity.
 \cite{wang-2009}
This coincidence
 implies that 
$\Delta \theta_K/\theta_K$
(or the identical pump--induced 
ellipticity)
reflects  fs dynamics of 
$\Delta S_z/S$, whose 
 sign correlates with 
the switchings of equilibrium magnetization 
induced by transverse $B$--field as discussed next.

$S_z$ varies smoothly with increasing 
or decreasing $B$--field
(Eq. \ref{anis-1}), 
consistent with the observed 
smooth variation of $\theta_K(B)$.\cite{wang-2009} 
In contrast, 
when the  sweeping of $B$--field 
 switches the magnetization 
between 
 $X^{\pm}$ and $Y^{\pm}$ 
equilibrium 
states, 
as seen in the static 
Hall magnetoresistivity, 
the  direction of  pump--induced 
fs  component  $\Delta S_z$ 
also switches sign. 
As discussed below, this 
behavior 
is consistent with our theory. 
The observed dependence
of the sign of fs  $\Delta \theta_K$ on the equilibrium easy axis
cannot be explained 
by invoking conventional nonlinear 
optical 
effects or  magnetization amplitude ``longitudinal'' changes.
\cite{beau,theory,demag,wang-rev} 
When the latter dominate, 
 $X^+$ ($X^-$) give the  {\em same} $\Delta S_z$
as $Y^+$ ($Y^-$). Indeed, here 
the two in--plane easy axes and, thereby, the 
in--plane magnetic states 
are equivalent (symmetric) 
with respect to
the probe propagation
 direction, which is  perpendicular to the X--Y plane. 
However, Fig.\ref{Fig3}(d) ($B$=0)  
and   Figs.\ref{Fig3}(c) and (e) 
($B$=$\pm 0.2T$)  
 clearly show that this is not the case.
This is  
in sharp contrast to large $B$=$\pm$1T: 
 Figs. \ref{Fig3}(b) and (f) show the same 
exact fs changes
for both increasing or decreasing $B$, which thus do not depend on easy axis. 
 As we discuss below, 
fs magnetization re--orientation 
due to fs spin--orbit torque 
 diminishes  
with increasing perpendicular $B$, 
consistent with the above behavior.
 
For $B$=0, Fig.\ref{Fig3}(d) 
 reveals 
a {\em symmetric and opposite} 
out--of--plane fs canting 
$\Delta S_z$(t)  between the 
 $X_0$ and $Y_0$ initial states.
 In this case, the initial 
magnetization ${\bf S}_0$ lies inside the sample plane
 and
 $S_z$$\approx$0 in equilibrium
(Fig. \ref{Fig3}(a)).
Thus, the 
observed 
 $\Delta S_z$(t) 
cannot be associated 
with an amplitude change, 
as it occurs in a direction 
[001]
perpendicular to 
${\bf S}_0$.
For large $B$, the magnetization aligns with the $B$--field  
along [001], 
$S_z$$\approx$$S$,
and $\Delta S_z$(t)  reflects longitudinal 
fs changes in  magnetization  amplitude. 
 \cite{theory,axt} 
For $B$=0, $S_z$$\approx$0
and  $\Delta S_z$(t) reflects 
transverse changes in magnetization direction. 
The opposite sign of laser--induced 
femtosecond $\Delta S_z$(t) 
between the $X^{\pm}_0$ and $Y^{\pm}_0$ 
states,  Fig.\ref{Fig3}(d),
can  only arise from  fs 
magnetization rotation 
towards opposite out--of--plane directions.  
Furthermore, 
except for the sign difference, 
 the  fs temporal profiles of 
$\Delta S_{z}/S$
in Fig.\ref{Fig3}(d)
are  {\em symmetric} 
between $X_0$ and $Y_0$.
This symmetry implies that the 
out--of--plane $\Delta S_z$ 
is driven by 
a laser--induced anisotropy 
field pulse
that 
points close to the diagonal direction
between $X_0$ and $Y_0$.
The step--like temporal profile implies that this 
  field 
has $\sim$100fs duration.
The above experimental 
observations are consistent 
with the direction and duration of 
$\Delta {\bf s}_h$  calculated in 
Fig. \ref{Fig2} 
for anisotropy parameter 
$\Delta_{pd}/\Delta_{so}$$\sim$1/3
similar to (Ga,Mn)As.
 Such   calculated   spin--pulse, 
discussed further below, 
exerts a fs spin--torque 
 $\propto \Delta {\bf s}_h \times {\bf S}_0$,   
whose out--of--plane direction 
changes sign 
while its magnitude remains the same 
for ${\bf S}_0$ along $X_0$ or $Y_0$. 
Note that laser--induced thermal effects 
due to spin--lattice coupling 
can change the equilibrium easy axis, \cite{nemec-2013,qi}
but such changes occur gradually in time over many ps. 
In contrast, here we observe 
step--like magnetization changes that follow the 
100fs laser pulse and are consistent 
with the predicted  fs spin--orbit torque.

\begin{figure}[t]
\begin{center}
\includegraphics[scale=0.38]{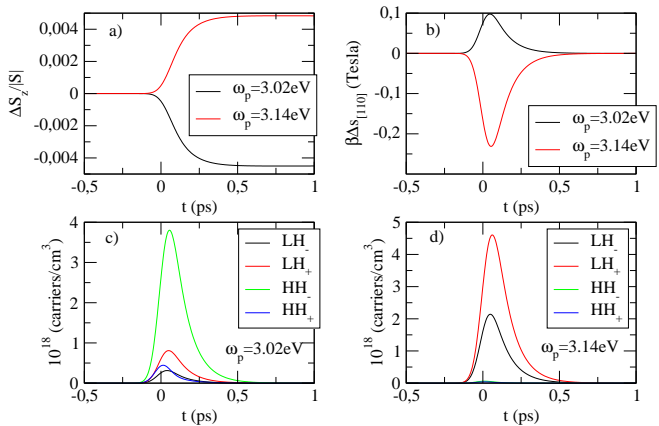}
\caption{(Color online) 
Frequency dependence of 
local and mobile 
spin dynamics and 
photohole populations 
following excitation by a 100fs linearly--polarized
laser pulse with low pump fluence 
$\sim$10$\mu J/cm^2$.
The 
initial magnetization points along the $X^+$ easy axis.  
(a): 
Comparison of ``sudden'' out--of--plane magnetization 
for $\hbar \omega_p$=3.14eV (LH  optical transitions) and 
 $\hbar \omega_p$=3.02eV (HH  optical transitions).
(b): Comparison of non--adiabatic 
photoexcited hole spin component along 
[110] for the two above
frequencies. 
(c): 
Coherent 
photoexcited hole populations 
of the  four  different exchange--split 
HH and LH bands 
for $\hbar \omega_p$=3.14eV.
(d): Same as (c) for 
 $\hbar \omega_p$=3.02eV. 
}
\label{Fig4}
\end{center}
\end{figure}

To relate our calculations to 
 Fig.\ref{Fig3},  
 we first take 
$B$=0 
and show in 
Fig. \ref{Fig4} 
the 
 spin and 
charge 
dynamics for 
 a  single linearly--polarized 
100fs laser pulse,  with 
$E_{0}$=2$\times
10^{5}$V/cm
similar to  the experiment.
The   initial state is  $X_0^+$.
We compare  the spin and  charge population 
 dynamics 
for two different laser
 frequencies,
  $\hbar \omega_p$=3.14eV 
and 
$\hbar \omega_p$=3.02eV, tuned   
to excite different  band continua  around 3.1eV.
In Fig. \ref{Fig4}(a)   
we show the development 
in time 
of the optically--induced 
out--of--plane local spin component  
$\Delta S_z$(t). 
The calculated step--like fs temporal 
profile agrees with  Fig. \ref{Fig3}.
Furthermore, 
we observe a reversal 
in  the direction of 
$\Delta S_z$
 when tuning the 
photoexcitation frequency.
The fs spin--orbit torque 
leading to such  $\Delta S_z$(t) 
is exerted  by the 
 photohole spin--pulse $\Delta {\bf s}_h$(t), 
whose component
along  the diagonal [110] direction is 
shown in Fig. \ref{Fig4}(b)
for the two above frequencies. 
 The magnitude, direction, and temporal profile 
of both local and mobile spin 
components in Figs. \ref{Fig4}(a) and (b) 
 are consistent with the 
experimental results of 
Fig. \ref{Fig3}(d).

Important for  controlling the 
four--state magnetic memory 
 is that we are able 
to reverse 
the direction 
of the
out--of--plane magnetization tilt $\Delta S_z$, 
Fig.\ref{Fig4}(a), and 
photoexcited  hole spin--pulse, 
Fig.\ref{Fig4}(b), 
by exciting $e$--HH ($\hbar \omega_p$=3.02eV)
or $e$--LH ($\hbar \omega_p$=3.14eV)
optical transitions.
The origin of this spin--reversal
can be seen 
by comparing the 
total  
populations 
$\frac{1}{V} \sum_{{\bf k}} 
\Delta \langle \hat{h}^{\dag}_{-{\bf k}n}\hat{h}_{-{\bf k}n}\rangle
$ 
for the four different 
exchange--split HH and LH valence bands $n$
in all \{111\} {\bf k}--directions
These band--resolved spin--polarized 
total  populations 
 are  shown in 
Figs. \ref{Fig4}(c) and (d)
as function of time
for $T_1$=100fs, which is 
comparable to the measured \cite{spinrelax} 
and calculated \cite{wu} hole spin 
relaxation time.
More than one  bands are populated simultaneously
due to the 
energy dispersion and  laser--pulse--width.
With frequency tuning 
we create a controlled short--lived 
imbalance 
between 
exchange--split bands 
with  different spin--orbit couplings
and 
spin admixtures. In this way 
we coherently  control the superposition of 
spin--$\uparrow$ and spin--$\downarrow$ 
states prior to spin  relaxation,  
here during the 100fs  pulse.
The 
order of magnitude of the 
photocarrier densities,
calculated 
by including the band continua 
along all eight \{111\}
{\bf k}--directions 
and using  the GaAs tight--binding parameters
of Ref.[\onlinecite{vogl}]
(Appendix \ref{cont}),
agrees 
with the 
experimentally--measured 
densities $n$$\sim$6$\times$10$^{18}$/cm$^3$
for the same pump fluence.
For such 
photohole populations, 
we also obtain 
 $\Delta S_z/S$ 
with same order of 
 magnitude and direction 
as  in the experiment 
(compare 
Figs. \ref{Fig4}(a)  and \ref{Fig3}(d)). 
The calculated $\sim$250mT 
 component of 
$\beta \Delta {\bf s}_h$(t)
along  [110], 
Fig. \ref{Fig4}(b), agrees with the 
100fs magnetic anisotropy 
field
 extracted from Fig. \ref{Fig3}(d)
and is larger than 
typical  fields 
obtained from 
calculations that assume a non--equilibrium 
Fermi--Dirac distribution.
 \cite{nemec-2013}
This theory--experiment 
agreement  indicates that 
non--thermal populations with 
 lifetimes $T_1$=100fs comparable to the 
hole spin lifetimes in (Ga,Mn)As \cite{wu,spinrelax} 
can explain the 
observed impulsive $\Delta S_z$(t).

\begin{figure}[t]
\begin{center}
\includegraphics[scale=0.36]{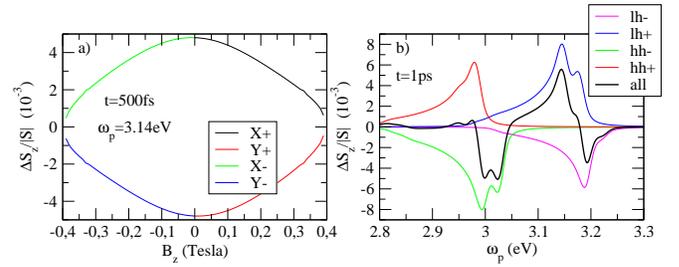}
\caption{
(Color online) The 
fs magnetic hysteresis and frequency dependence 
of photoinduced magnetization canting $\Delta S_z/S$
characterize the proposed fs spin--orbit torque non--adiabatic 
mechanism. 
(a): The direction  of out--of--plane 
fs component 
$\Delta S_z/S$ depends on the easy axis (magnetic hysteresis) 
and magnetic 
field direction. 
This magnetic hysteresis diminishes with increasing perpendicular 
$B$--field due to suppression of laser--induced 
magnetization re--orientation, which separates
experimentally ``transverse'' 
and ``longitudinal'' spin dynamics.   
(b): Frequency dependence of  
laser--induced 
$\Delta S_z/S$ 
and its  individual contributions 
from  the four exchange--split HH and LH 
bands, 
calculated at  $t$=1ps
for $E_0$=2$\times$10$^5$V/cm. 
Initial magnetization points along $X^+$.
The band continua significantly affect the frequency 
dependence of $\Delta S_z$(t)  
as compared to discrete--${\bf k}$ special point 
calculations. 
}
\label{Fig5}
\end{center}
\end{figure}

Further evidence in favor of the proposed 
fs spin--orbit torque mechanism 
 is obtained from 
the pump--induced fs magnetic hysteresis 
observed in the experiment of Fig. \ref{Fig3}.  
In Fig. \ref{Fig5}(a) we compare the fs  canting 
$\Delta S_z$/S,  calculated  at  $t$=500fs 
as function of $B$ 
 along [001],
for 
the four 
$B$--dependent 
magnetic states  $X^{\pm}$ and $Y^{\pm}$.
Fig. \ref{Fig5}(a) shows that  
switching between $X$ and $Y$  states, 
induced in the experiment by sweeping  $B$, 
switches the sign of $\Delta S_z$(t) 
(fs magnetic hysteresis). 
Furthermore, 
Fig. \ref{Fig5}(a) shows that
fs magnetization re--orientation 
diminishes with increasing $B$ and  practically
vanishes  at high magnetic fields. 
The above results,
consistent with 
Fig. \ref{Fig3}, explain the observed coincidence 
of  $\Delta S_z$  switchings 
with static planar Hall effect 
switchings \cite{wang-2009}
and the absence of hysteresis at high $B$. 
While nonlinear effects such as 
dichroic bleaching  contribute to the  fs 
magneto--optical signal, the observed systematic 
$B$--field dependence and 
magnetic hysteresis in the sign of 
$\Delta \theta_K/\theta_K$
indicate non--adiabatic origin consistent with our calculations
of fs spin--orbit torque.

For high B--fields, only longitudinal  changes 
in magnetization amplitude \cite{demag} and 
nonlinear optical effects  
contribute to Fig. \ref{Fig3}(b) and Fig. \ref{Fig3}(f).
The mean--field 
density matrix factorization 
 used here does not capture magnetization amplitude changes, 
which  appear
 at the level of electron--magnon 
spatial 
correlations.\cite{axt,kapet-corr}  
As discussed in Ref.[\onlinecite{theory}], 
any photoinduced imbalance 
of spin--$\uparrow$ and spin--$\downarrow$ 
states will lead to 
fs demagnetization and inverse Overhauser effect, 
which is independent of easy axis direction. 
While such imbalance may arise from photoinduced 
changes in the Fermi--Dirac temperature and chemical potential, 
a large  
electronic temperature increase  
is required to produce the 
broad distributions implied 
by the magnitude of the 
experimentally--observed effects. \cite{theory}
The broad non--thermal populations considered here 
create an imbalance that, 
 for  $T_1 \le$100fs,
follows 
the 
laser pulse. In contrast, relaxation to the lattice 
bath is  slower (ps). \cite{theory}
Both ``longitudinal'' and ``transverse''
fs spin  dynamics 
arise from the competition 
of spin--orbit and magnetic--exchange interactions, 
but manifest themselves 
differently for different photoexcitation conditions 
and external magnetic fields. 
Femtosecond  demagnetization (decrease in Mn spin amplitude) 
through dynamical polarization of longitudinal hole spins
dominates for 
high fluences of 100s of 
$\mu$J/cm$^{2}$. 
\cite{spinrelax}
Our proposed pulse--train 
scheme  achieves spin rotational switching
by using lower pump intensities that reduce fs demagnetization.

As already seen in 
Fig. \ref{Fig4}, 
by coherently controlling the 
non--thermal  population imbalance between the 
four exchange--split HH and LH bands,
we can control the  
direction 
of out--of--plane $\Delta S_z$/S. 
This is seen more 
clearly in 
Fig. \ref{Fig5}(b), which 
shows the frequency--dependence of  $\Delta S_z$/S 
and  compares it to
the 
 individual  contributions  
 obtained 
by retaining one valence  band 
at a time. The non--equilibrium population  of 
band states 
with different spin admixtures 
leads to different directions of laser--induced 
$\Delta  S_z$(t).  For example, 
photoholes excited in the two  exchange--split 
(HH or LH) 
valence bands induce opposite 
out--of--plane tilts.
The finite pulse--duration and strong 
band dispersion and 
non--parabolicity  (Appendix \ref{cont} and Fig. \ref{Fig1}(a)) 
lead to  different 
populations of more than one bands
at all frequencies.
We conclude that optical 
control 
of the 
photoexcited carrier populations
can 
be used to 
 switch the  directions of photoexcited  
fs spin--orbit torques and, in this way, control 
the direction of fs magnetization  tilts.

The precise magnitude of the proposed effects
depends on the relaxation  timescales. 
The non--thermal populations are
created during the 100fs laser pulse, 
via  $e$--$h$ optical polarization,
with dephasing time $T_2$,  driven 
by the laser field.
 They  
relax  on a timescale $T_1$ that  is comparable to 
hole spin relaxation. \cite{spinrelax,wu}
The above characteristic 
times are expected to be in the 10-200fs 
range. \cite{spinrelax,wu}
For pump fluences of $\sim$10$\mu J/cm^2$, 
the experiment 
gives $\Delta S_z/S$$\sim$0.5\%,
which we reproduced for
$T_1$=100fs and $T_2$=50fs.
This  contribution 
decreases 
 to $\Delta S_z/S$$\sim$0.01\%  
as $T_2$ decreases to 
3fs with fixed $T_1$=100fs. 
For fixed short $T_2$=10fs, 
$\Delta S_z/S$ varies between $\sim$0.05-0.1\% 
as $T_1$ varies between 30fs and 100fs. 
We conclude that 
the fs spin--orbit torque 
contribution 
has the same order of magntitude  as the 
experimental result 
unless both $T_1$ and $T_2$ are only  few fs. 
From now on we fix $T_1$=100fs and $T_2$=50fs.

\begin{figure}[t]
\begin{center}
\includegraphics[scale=0.36]{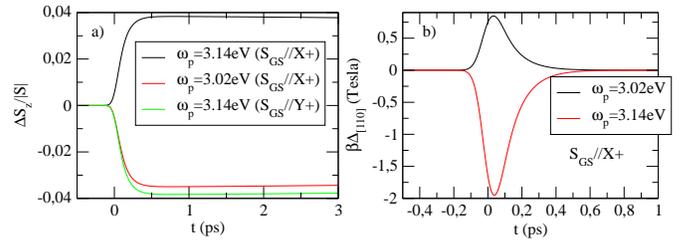}
\caption{
(Color online) Calculated fs spin dynamics 
similar to Fig. \ref{Fig4} 
but with order of magnitude higher 
pump fluence 
$\sim$100$\mu J/cm^2$.
(a) Comparison of out--of--plane magnetization 
components 
for two 
different initial magnetic states and $\omega_p$.
(b) Photohole fs anisotropy 
fields along [110] 
for the two  $\omega_p$. 
}
\label{Fig6}
\end{center}
\end{figure}

The non--thermal/non--adiabatic fs spin--orbit torque 
 contribution 
can be enhanced by increasing the laser intensity. 
Fig. \ref{Fig6}(a) shows that, 
for easily attainable 
$\sim$100$\mu J/cm^2$
 pump fluences, \cite{nemec-2013} 
 the ``sudden'' magnetization 
tilt 
increases to 
 $\Delta S_z/S$$\sim$4\%
(for $E_0$=7$\times
 10^{5}$V/cm), 
while  
Fig. \ref{Fig6}(b) 
shows that
$\beta \Delta {\bf s}_h$(t) 
along [110] 
then grows into the Tesla range. 
The precise value is sample--dependent  
and  depends on the 
relaxation.
While the quasi--equilibrium  contribution ${\bf H}_{FS}$
is limited by the 
chemical potential, 
$\beta \Delta {\bf s}_h$  is controlled by the laser frequency. 
The different intensity--dependence and temporal profiles 
of thermal and coherent/non--thermal carrier 
spin components can separate these two  contributions.
A distinct impulsive component of 
fast magnetic anisotropy 
 was 
observed 
in the ps magnetization 
trajectory measured in Ref.[\onlinecite{nemec-2013}] 
for pump fluences above $\sim$70$\mu J/cm^2$
and $\hbar \omega_p$$\sim$1.5eV. 
Finally, Fig. \ref{Fig6}(a)
compares the dynamics for 
initial magnetization along the $X_0^+$ or $Y_0^+$ 
 easy axis for $B$=0.  
Similar to the experiment of  
Fig. \ref{Fig3}(d),  
it displays 
symmetric temporal profiles
of  $\Delta S_z$(t), 
with opposite signs 
for the two perpendicular easy axes.
In this way, we can 
distinguish the  magnetic states
within 100fs.
The equal magnitude of $\Delta S_z$ 
between the two perpendicular in--plane easy axes arises 
from the 
diagonal direction of 
$\Delta {\bf s}_h$ 
obtained  
for  $\Delta_{pd}/\Delta_{so}$$\sim$1/3 
as in (Ga,Mn)As (Fig. \ref{Fig2}).
The overall agreement 
between theory and experiment 
suggest  that
a
magnetic state can be read within 100fs,
by monitoring the  direction of 
 out--of--plane laser--induced 
magnetization canting.
In the next section we discuss how one could 
also switch  the four--state 
magnetic memory with low intensity by using pulse--shaping. 

\section{Initiating  deterministic 
switchings 
with a laser--pulse--train}
\label{one-train} 

\begin{figure}[t]
\begin{center} 
\includegraphics[scale=0.5]{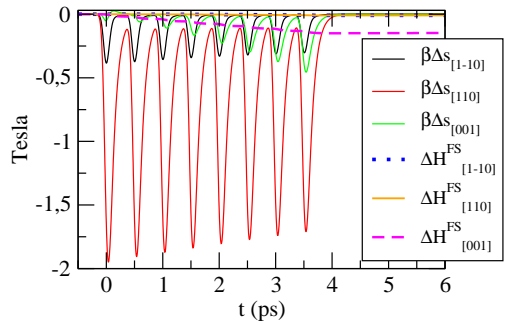} 
\caption{(Color online)
Comparison of 
non--thermal and quasi--thermal components of 
laser--induced 
magnetic anisotropy fields 
$\beta \Delta {\bf s}_h$(t) 
and $\Delta {\bf H}_{FS}$(t) 
during coherent nonlinear photoexcitation 
with a train of $N$=8
100fs  laser pulses
separated by 500fs, 
 with 
$E_{0}$=7$\times
 10^{5}$V/cm and $\hbar \omega_p$=3.14eV. 
 } 
\label{Fig7}
\end{center} 
\end{figure}

A single 100fs laser pulse 
 with $\sim$10--100$\mu$J/cm$^2$ fluence 
excites magnon oscillations
around the equilibrium  easy axis.
Laser--induced 
switching of the magnetization to a different  magnetic 
state 
requires  photoexcitation of  a stronger 
``initial condition'' 
$\Delta {\bf S}$(t).
While switching may be possible by increasing 
the pump intensity, \cite{asta} 
 pulse--shaping \cite{pulseshaping} 
can be used to initiate it 
in a 
more controlled way 
while keeping the laser fluence per pulse as low as possible
to reduce fs
electronic heating of spins.
We 
coherently control 
$\Delta {\bf s}_h$(t)
with 
$M$
time--delayed  trains 
of $N$  Gaussian $\tau_{p}$=100fs laser pulses:
\begin{eqnarray}
 {\bf E}(t) &=& \sum_{j=1}^M  {\bf E}_{0} 
 \sum_{i=1}^N
\exp[-(t- \tau_{j} 
-  \Delta \tau_{ij}) 
^{2}/\tau_{p}^{2}] 
\times \nonumber \\
&& \exp[-i\omega_{p}^{(j)} (t-\tau_{j}
-\Delta \tau_{ij} 
)].
\label{eq:train}
\end{eqnarray}
We tune 
$\tau_{j}$, the time delay of the $j$-th 
laser--pulse--train, and 
$\omega_{p}^{(j)}$, the pulse--train  central 
frequency but fix $\Delta \tau_{ij}$=500fs
for simplicity. 
In this section we consider $M$=1
and 
 control the net duration of successive  
 fs spin--orbit torques 
by using a train 
of $N$ pulses.
In Fig. \ref{Fig7} we compare the  
components of 
$\beta \Delta {\bf s}_h$(t) and $\gamma \Delta {\bf H}_{FS}$
in the
coordinate system  
 defined by the  
[110], [1-10], and [001]
directions. 
We use the same  $\sim$100$\mu$J/cm$^{2}$ fluence
as in 
Fig. \ref{Fig6}, 
but 
increase the number of pulses 
from $N$=1 to $N$=8.  
The non--thermal contribution 
$\beta \Delta {\bf s}_h$(t) 
now prevails, while 
$\Delta {\bf H}_{FS}$(t) 
 builds--up 
as $\Delta {\bf s}_h$ drives $\Delta {\bf S}$(t) 
and forces the spin of the 
Fermi sea bath to 
adjust to the new 
direction of ${\bf S}$(t).\cite{chovan} 
$\Delta {\bf S}$(t)
 builds--up
in a 
 step--by--step fashion
 well before relaxation,  
driven by photoexcited sequences of successive fs  spin--orbit 
torques.

\begin{figure*} 
 \includegraphics[scale=0.23]{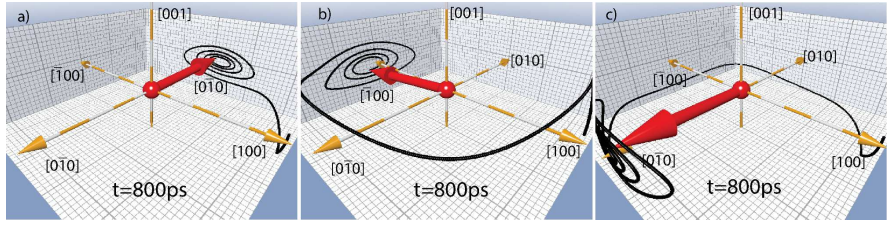}
\caption{
(Color online) 
Magnetization switching trajectories
 from $X^+$ to
the other three magnetic states,  
 triggered  
by trains with increasing number of laser pulses
$N$ and $E_{0}$=7$\times
 10^{5}$V/cm.
All switchings are followed by pronounced magnetic ringing.
 (a): Counter--clockwise 
90$^{o}$ switching  
$X^{+}$$\rightarrow$$Y^{+}$, initiated  by HH 
photoexcitation with $N$=7 pulses.
(b): 180$^{o}$ magnetization 
reversal via 
clockwise 
pathway 
$X^{+}$$\rightarrow$$Y^{-}$$\rightarrow$$X^-$,  
initiated by LH
photoexcitation with $N$=9 pulses.
(c): Photoexcitation as in (a), but with 
$N$=12 pulses. By increasing $N$
we can move the magnetization past the 
$Y^{+}$ and $X^{-}$ 
intermediate states and access the 
$Y^{-}$ state via the 
270$^{o}$ counter--clockwise pathway 
$X^{+}$$\rightarrow$$Y^{+}$$\rightarrow$$X^-$$\rightarrow$$Y^-$.
}  
\label{Fig8}
\end{figure*}

 $\Delta {\bf H}_{FS}$(t)
originates  from the 
spin of the thermal hole Fermi sea 
and is therefore  restricted 
 by the 
Fermi--Dirac distribution. Such populations
give anisotropy 
fields of  the order of few 10's of mT
in (Ga,Mn)As, \cite{nemec-2013,jung} 
restricted by the  $\sim$$\mu$eV free energy differences
close to the Fermi level.  
On the other hand, the 
experiments observe 
anisotropy fields that are at least one 
order of magnitude higher.
\cite{nemec-2013,wang-2009}
Fig. \ref{Fig7} 
 shows 
  $\beta \Delta {\bf s}_h$(t) 
calculated in the time--domain, 
by solving density matrix equations of motion 
after taking into account the (Ga,Mn)As 
bandstructure.
It comes from the short--lived non--thermal populations 
photoexcited at $\hbar \omega_p$=3.14eV. 
These populations are not restricted by the 
Fermi--Dirac distribution
and access parts of the Brillouin zone that 
are empty close to quasi--equilibrium.  
 $\beta \Delta {\bf s}_h$(t)  can grow to 
$\sim$2T along [110] 
for experimentally--relevant pump fluences 
and  photocarrier 
thermalization times $T_1$$\sim$100fs.
For such fast relaxation, $\Delta {\bf s}_h$(t)
follows the laser--pulse--train 
temporal profile and  
the relative phase 
of  consequative pulses does not play a role. 
However, 
$\Delta {\bf s}_h$(t) 
is not the same 
for different  pulses, as 
the 
 non--equilibrium electronic states 
change non--adiabatically with $\Delta {\bf S}$(t)
(Appendix \ref{nonadia}).

We now show that, by tuning $N$,  we can initiate switching
rotation 
to any one of the available  magnetic states,
 instead of  exciting 
magnon oscillations around the equilibrium easy axis. 
Fig. \ref{Fig8} shows three  magnetization 
switching trajectories up to long times 
$t$=800ps. These ps
trajectories  
are initiated  at $t$=0
 by $N$=7 
(Fig. \ref{Fig8}(a)), $N$=9 
(Fig. \ref{Fig8}(b)), or $N$=12 
(Fig. \ref{Fig8}(c)) laser pulses
with $\sim$100$\mu$J/cm$^{2}$ 
fluence.
By increasing $N$, 
we switch from $X^+$ to
all three of the other 
magnetic states 
$Y^+$, $X^{-}$, and $Y^{-}$. 
In Fig. \ref{Fig8}(a), 
$N$=7 pulses with  $\hbar \omega_p$=3.02eV
(HH photoexcitation)  
initiate a 
counter--clockwise 
90$^{o}$ 
switching 
rotation 
that  stops  after reaching  the next magnetic state, $Y^+$, 
within  $\sim$80ps.
The magnetization 
oscillates around the final state 
with a significant amplitude 
that cannot be controlled 
with a single pulse--train 
(magnetic ringing).\cite{stohr} 
This  ringing results from the 
weak (nanosecond) Gilbert damping of 
the local--spin precession 
observed 
 in annealed 
(Ga,Mn)As. \cite{qi,nemec-2013}
While
magnetic ringing 
can make the 
multiple 90$^{o}$ 
switchings
unstable, 
below we show that
we can  
suppress it   
 by exerting
 opposing fs spin--orbit torques. 
By increasing the number of  pulses 
to $N$=9, the magnetization  continues
past $Y^+$  
to the next available state, $X^{-}$.
 Fig. \ref{Fig8}(b)   shows  a
 magnetization reversal
via clockwise instead of counter--clockwise 
rotation, since 
 $\hbar \omega_p$=3.14eV
  excites $e$--LH instead of 
$e$--HH optical transitions. 
 This $X^+$$\rightarrow$$Y^-$$\rightarrow$$X^-$ 
pathway 
completes within 
$\sim$150ps in (Ga,Mn)As and is again   
 followed by  magnetic ringing.
By increasing the number of pulses 
further, 
to 
$N$=12, the initial fs spin--orbit torque  
is sufficient
to move the  magnetization even beyond $X^{-}$. 
 Fig. \ref{Fig8}(c) shows 
270$^{o}$ switching to the $Y^{-}$  state
 within $\sim$200ps, following the 
$X^+$$\rightarrow$$Y^+$$\rightarrow$$X^-$$\rightarrow$$Y^-$ 
 pathway initiated by $e$--HH photoexcitation.

\section{Optical control of 
sequential 90$^{o}$ switchings 
between   four  states} 
\label{Protocols} 

\begin{figure}[t]
\begin{center} 
\includegraphics[scale=0.36]{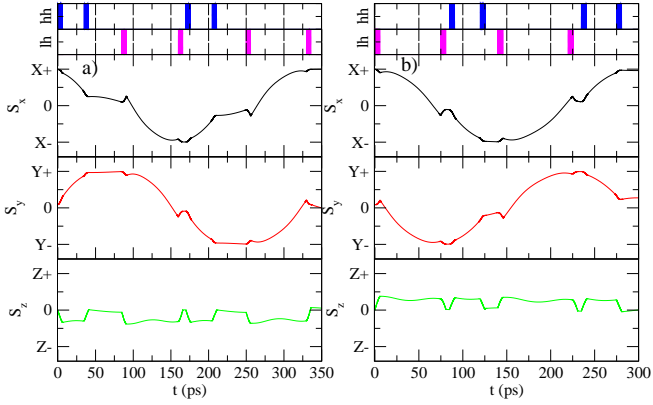}
\caption{ (Color online) 
Two protocols for  360$^{o}$ control of  the full four--state 
magnetic memory via four sequential 90$^{o}$ switchings
that  stop and  restart at 
each intermediate magnetic state. 
 (a) Counter--clockwise 
 sequence 
$X^{+}$$\rightarrow$$Y^{+}$$\rightarrow$$X^{-}$$\rightarrow$$Y^{-}$$\rightarrow$$X^{+}$, 
(b) Clockwise 
sequence
 $X^{+}$$\rightarrow$$Y^{-}$$\rightarrow$$X^{-}$$\rightarrow$$Y^{+}$$
 \rightarrow$$X^{+}$.  
Upper panel: Laser--pulse--train ($N$=12) timing sequences 
and colors.
 Blue pulses excite  
HH optical transitions,  
magenta pulses excite LH transitions. 
$E_{0}$=7$\times
 10^{5}$V/cm
corresponds to pump fluence 
$\approx$100$\mu J/cm^2$. 
}  
\label{Fig9}
\end{center} 
\end{figure}

In this section we 
 study the possibility to 
manipulate fs spin--orbit torques 
in order 
to gain full ultrafast access of this  four--state 
magnetic memory. 
Fig. \ref{Fig9} shows 
two switching protocols 
that achieve 
 360$^{o}$ 
control of the
magnetic states of Fig. \ref{Fig1}(b).
The upper panel shows the 
sequences of 
laser--pulse--trains used to 
control the 
four  sequential 90$^{o}$ switchings.  
 Two different laser frequencies   excite
$e$--HH or $e$--LH optical transitions
that stop and restart the magnetization motion 
at each of the four magnetic states.
By tuning the laser frequency,  we 
choose the direction of 
this
multi--step
 switching process, which 
takes place via   counter--clockwise (Fig.\ref{Fig9}(a)) 
or  clockwise 
(Fig.\ref{Fig9}(b)) magnetization 
rotations forced to stop at all intermediate states.
To  control the photoexcited 
 $\Delta {\bf s}_h$(t) and fs spin--orbit torques, 
  we turn three experimentally accessible  
``knobs'': 
(i) {\em Pulse--shaping} 
\cite{pulseshaping}
by changing  $N$, which  
controls 
the net duration and temporal profile of the 
spin--orbit torques.
In this way, we tailor 
$\Delta {\bf S}$(t)
that initiates or modifies 
the switching rotations 
 with low intensity per laser pulse.
(ii)
{\em Frequency--tuning} 
enables  selective 
photoexcitation of exchange--split 
LH or HH 
non--equilibrium populations 
with different 
superpositions of
spin--$\uparrow$ and 
spin--$\downarrow$ 
 states. 
In this way we control the 
population  imbalance 
that 
decides  the  directions of 
$\Delta {\bf s}_h$, fs spin--orbit torque, 
and 
  $\Delta {\bf S}$(t).
(iii) By controlling the
{\em time--delays} $\tau_j$,  
one can exert fs spin--orbit torques 
at  desirable   times 
in order to   stop and restart the  switching 
rotation at all  intermediate states
and suppress  magnetic ringing.
This is discussed further in the next section.  
To understand the role of the 
twelve 
 laser--pulse--trains
in Fig. \ref{Fig9}, 
we note the following points: 
(i) a laser--pulse--train initiates switchings or 
magnon oscillations 
via fs
spin--orbit torque 
with direction   that depends on {\em both} laser 
frequency
and  magnetic state, 
(ii)  when  the magnetization reaches a new 
magnetic state,  
we use a
laser--pulse--train 
to exert opposing fs
spin--orbit--torques, 
in a direction that 
stops the switching 
rotation  and 
suppresses the magnetic ringing
so that we can access the  state, and (iii)  
when we are ready to move on, 
a
laser--pulse--train with the appropriate 
color  
restarts the 360$^{o}$ switching process
by exerting 
fs spin--orbit torques 
in the desirable direction.

Fig. \ref{Fig9}  
shows four  sequential  90$^{o}$
switchings 
controlled by 
$\Delta {\bf s}_h$(t).
In Fig. \ref{Fig9}(a), 
a counter--clockwise  $X^{+}$$\rightarrow$$Y^{-}$ 
switching
is initiated by $e$--HH 
photoexcitations
with $N$=12 pulses.
After  $\tau$=35ps, 
the magnetization reaches the vicinity of  
the intermediate $Y^+$ state. 
We then 
stop the
switching process by 
exciting 
$e$--HH  optical  transitions.
We restart the motion 
at $\tau$=85ps,
after waiting for about 50ps, 
by using $e$--LH  photoexcitations 
to switch the magnetization to 
the  $X^{-}$ state. There we again stop the process
at $\tau$=160ps, by exciting $e$--LH  
optical transitions.  
We restart 
at $\tau$=170ps with $e$--HH  photoexcitations, which 
trigger switching
 to  $Y^-$. This switching completes within $\sim$35ps, 
after we stop the motion with $e$--HH photoexcitations at
  $\tau$=205ps. 
We finish the  360$^{o}$ switching loop by
using
$e$--LH photoexcitations
to restart the  counter-clockwise motion 
back to 
$X^+$,
 at 
 $\tau$=250ps, 
 and later 
to terminate the process  at $\tau$=330ps.  
 Fig.\ref{Fig9}(b)
 shows an opposite 
clockwise switching sequence
$X^+$$\rightarrow$$Y^-$$\rightarrow$$X^-$$\rightarrow$$Y^+$$\rightarrow$$X^+$, obtained 
by changing the laser--pulse frequencies  
from $e$--HH to $e$--LH excitations and vice--versa. 
In this case,
$e$--LH optical transitions 
with $N$=12 pulses 
trigger clockwise magnetization rotation,
 which we
 suppress at $Y^-$ with LH excitations at  $\tau$=75ps.
We restart the process with $e$--HH photoexcitation at $\tau$=85ps 
and suppress 
it again at $X^-$ with $e$--HH optical transitions
 at $\tau$=120ps.
We restart with $e$--LH excitation at 
$\tau$=140ps and 
 switch to $Y^{+}$, where we suppress the motion at 
$\tau$=225ps  with $e$--LH optical transitions. We complete a 
closed switching
 loop to the initial $X^{+}$ state with $e$--HH photoexcitation at 
$\tau$=235ps  
and suppress the rotation 
with $e$--HH optical transitions at $\tau$=275ps.
In the next section 
we analyze how tunable fs spin--orbit torque direction offers 
more flexibility for controlling switching rotations and 
magnetic ringing.

\section{Controlling  magnetic switching 
 and ringing 
with a laser--pulse--train}
\label{two-train} 

A  multi--state memory 
allows for  more elaborate  tests 
of optical  control schemes as compared to simply flipping 
the spin between two states.  
In order to selectively access  
 four different  magnetic states
via 360$^{o}$  
magnetization rotation,
we must be able to not only  initiate switchings, as in  
Fig. \ref{Fig8},  but also 
to suppress  the magnetization  
motion at any intermediate state. 
We must also be able to suppress the ringing due to 
 magnetization oscillations 
around an easy axis. 
Magnetic ringing arises from 
the weak damping of the magnetization precession
following excitation with either optical   or  
magnetic field pulses  
and limits the read/write times in 
many magnetic materials. \cite{stohr} 
One known way to reduce it is to 
take advantage of the phase 
$\Omega \tau$ 
of 
magnetization precession
with frequency  $\Omega$.
 \cite{stohr,Hashimoto-apl}
With  magnetic field pulses, 
this can be  done by adjusting the duration of the 
long pulse to the 
precession period.\cite{stohr} 
With ultrashort  laser pulses, 
one can suppress (enhance) the 
precession 
by exciting  when
$\Omega  \tau$=$\pi$ 
($\Omega  \tau$=2$\pi$),  
in  the same  way as 
at  $\tau$=0.
 \cite{Hashimoto-apl}
Such  coherent 
control of spin precession
is possible for 
 harmonic oscillations. 
Below we show that 
we can optically control 
both magnon oscillations  and  nonlinear switching rotations,
at all time
delays,
by applying either 
 clockwise or counter--clockwise 
fs spin--orbit torques  as needed.

\begin{figure}[t]
\centering  
\includegraphics[scale=0.35]{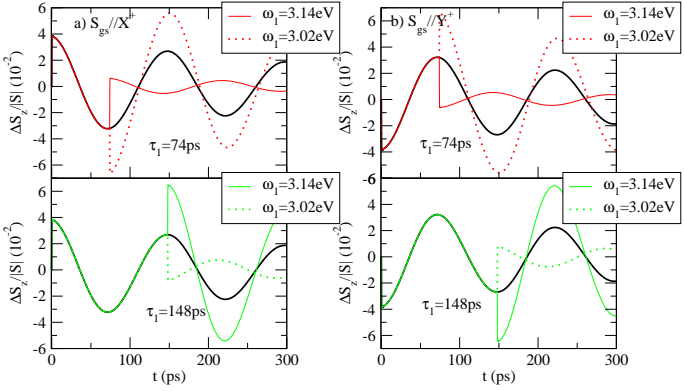}
\caption{(Color Online) 
Two 100fs laser pulses, delayed by $\tau$, control   
magnon oscillations via 
fs spin--orbit torque.
The first   pulse, $\hbar \omega_{p}$=3.14eV,  
starts the  precession (frequency 
$\Omega$) at $\tau$=0. 
 The second pulse, 
$\hbar \omega_{p}=3.02$eV 
or 
$\hbar \omega_{p}=3.14$eV,
arrives at
$\tau$=74ps ($\Omega \tau$=$\pi$) or 
$\tau$=148ps ($\Omega \tau$=2$\pi$).
Equilibrium magnetic state:
(a): $X^+$ and   (b): $Y^+$.
}
\label{Fig10}
\end{figure}

We start with the harmonic limit and 
demonstrate magnon control 
via fs spin--orbit torque with tunable direction.  
First we excite
at $\tau$=0
 magnon oscillations with frequency $\Omega$
(thick solid line in Fig. \ref{Fig10}). 
We thus initiate magnetization precession   
around the $X^+$ (Fig. \ref{Fig10}(a)) 
or the $Y^+$ (Fig. \ref{Fig10}(b))
easy axis  with 
$e$--LH   excitation
($\hbar \omega_p$=3.14eV).  
An impulsive magnetization 
at $\tau$=0
 is  observed in the 
ps  trajectory
of Fig. \ref{Fig10}.   
The initial phase 
of these magnon oscillations is opposite 
between the $X_0^+$ and $Y_0^+$  states,
due to the opposite 
directions of the fs spin--orbit torques
(Fig. \ref{Fig6}(a)).
We then send  a  control laser pulse
at $\tau$=74ps ($\Omega \tau$=$\pi$) or  at
$\tau$=148ps ($\Omega \tau$=2$\pi$), 
but use
either 
$\hbar \omega_p$=3.14eV ($e$--LH optical transitions) or 
$\hbar \omega_p$=3.02eV ($e$--HH optical transitions). 
By controlling the direction of fs spin--orbit torque
with such frequency tuning,  
 we show 
that we can 
both enhance and  suppress 
the amplitude of the magnetization precession at {\em both} 
$\Omega \tau$=$\pi$  and 
$\Omega \tau$=2$\pi$.   
While for 
$\Omega \tau$=$\pi$ we suppress magnetic ringing 
when applying the same fs spin--orbit torque as for 
$\tau$=0 ($\hbar \omega_p$=3.14eV), 
we  enhance it 
by applying an  opposite 
fs spin--orbit torque 
($\hbar \omega_p$=3.02eV). 
Similarly,
at time $\Omega \tau$=2$\pi$, 
we  enhance the ringing 
when applying fs 
spin--orbit torque in the 
same  direction  as for $\tau$=0 and  
suppress it 
by reversing the     direction.
We thus gain  
flexibility for controlling  magnon oscillations.


\begin{figure*} 
\centering  

\includegraphics[scale=0.5]{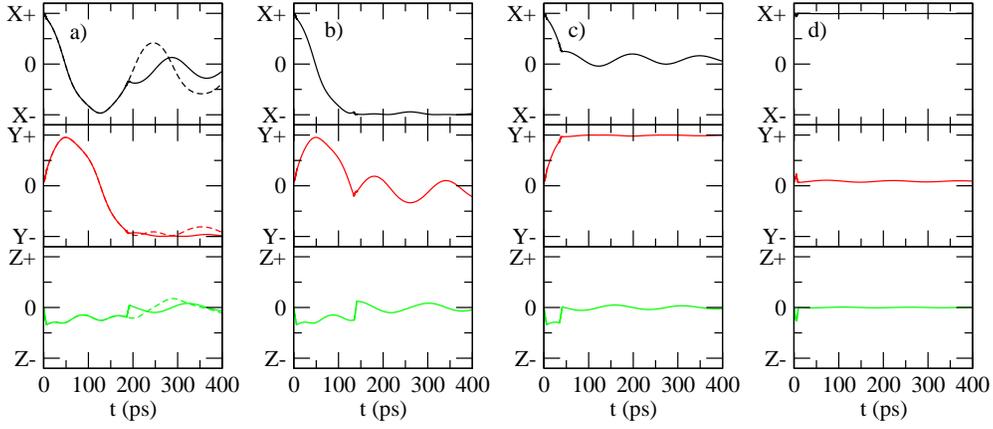}
\caption{(Color online) Time--dependence 
of  magnetization components 
controlled by a time--delayed fs spin--orbit torque
pulse train. 
(a):
A
 $X^+$$\rightarrow$$Y^+$$\rightarrow$$X^-$$\rightarrow$$Y^-$
switching pathway is initiated at $\tau$=0 with HH photoexcitation
(dashed line). After it completes,
the magnetic ringing is reduced  with a control laser--pulse--train 
 that exerts opposing fs spin--orbit torques  at any
 time
(solid line). 
(b): 
$X^{+}$$\rightarrow$$Y^{-}$ 
switching of (a) is terminated 
via opposing  fs spin--orbit torques 
after magnetization reversal to  $X^{-}$.
(c):
$X^{+}$$\rightarrow$$Y^{-}$ 
switching of (a) is terminated 
 via a  control laser--pulse--train 
after 90$^{o}$ 
rotation 
to  $Y^{+}$. 
(d): 
$X^{+}$$\rightarrow$$Y^{-}$ 
switching of (a) is  stopped 
immediately after it is initiated, 
via  opposing fs spin--orbit torque at $\tau$=2ps. 
}
\label{Fig11}
\end{figure*}

Unlike for harmonic  precession, 
switching also
involves  nonlinearities and anharmonic 
effects. 
In Fig. \ref{Fig11}(a), 
a 
$X^+$$\rightarrow$$Y^{+}$$\rightarrow$$X^{-}$$\rightarrow$$Y^{-}$ 
switching pathway (dashed line) 
is initiated at $\tau$=0 
as in Fig. \ref{Fig8}(c).  
  After about 
200ps, 
the magnetization switches to  $Y^{-}$,
after overcoming the 
intermediate states 
$Y^+$ and $X^-$.
The X--component of the magnetization then 
oscillates with a significant amplitude (magnetic ringing, 
dashed curve
in Fig. \ref{Fig11}(a)
and Fig. \ref{Fig8}(c)).
Fig. \ref{Fig11}(a) 
(solid curve)
demonstrates suppression of this ringing by a 
control laser--pulse--train that arrives 
at any  time after the switching is completed.  
To accomplish this,  we  
tune the direction 
and strength of
fs spin--orbit torques.
Figs. \ref{Fig11}(b) and (c) demonstrate that 
the control pulse--train 
can also stop the 
$X^+$$\rightarrow$$Y^{+}$$\rightarrow$$X^{-}$$\rightarrow$$Y^{-}$
switching 
at one of the 
intermediate magnetic states
before reaching $Y^-$. 
However, 
 we must use  different 
$\omega_p$ at 
 $Y^{+}$
and 
$X^{-}$
in order to 
get an opposing fs spin--orbit torque, 
as the direction of the latter depends on the magnetic state.  
In Fig. \ref{Fig11}(b) 
we  stop the switching at 
the
  $X^{-}$ magnetic state, 
after passing through Y$^{+}$, 
by exciting with 
$\hbar \omega_p$=3.14eV 
at $\tau$$\sim$100ps ($e$--LH photoexcitation).
Fig. \ref{Fig11}(c) shows that
we can stop 
at $Y^+$ after $\sim$35ps, 
by exerting  a clockwise spin--torque using  
$\hbar \omega_p$=3.02eV 
(HH photoholes).
A more dramatic demonstration 
of 
the flexibility offered by 
 fs  spin--orbit torque 
 is given in
Fig. \ref{Fig11}(d).
Here 
we initiate the 
$X^+$$\rightarrow$$Y^{-}$ switching 
 as above and then stop it immediately, by applying a 
control laser--pulse--train 
at 
$\tau$=2ps, 
i.e. long before any oscillations can develop. 
Instead of relying on the precession phase
as in 
 Fig. \ref{Fig10}, 
we apply a sufficiently strong clockwise fs spin--orbit torque 
that opposes the 
magnetization motion. 
In this way, we stop the magnetization 
at its tracks, 
after  a minimal motion without  oscillations. 
We conclude that coherent optical control of 
the mobile spin excited during fs laser pulses 
allows us to suppress both magnetic ringing and 
nonlinear switching rotations,  by
controlling the direction, duration, and magnitude 
of fs spin--orbit torques.

\section{Conclusions and Outlook}

In 
this paper we 
used 
density--matrix equations of motion 
with bandstructure
to  
describe photoexcitation and control of 
fs spin--orbit torques 
analogous to the static 
current--induced ones 
in spintronics.
In this all--optical way, 
we can 
initiate and 
control 
 multiple  magnetic  switchings and  magnetic ringing.
The proposed non--adiabatic  mechanism 
involves 
optical 
control of direction, 
magnitude, and temporal profile  of fs spin--orbit torque
sequences. 
This is achieved by 
tuning, 
via the optical field,  a short--lived population and spin
imbalance between exchange--split  bands 
with different spin--orbit interactions.
The photoexcited  spin magnitude and direction
depend on  symmetry--breaking 
 arising from the 
non--perturbative  competition 
of 
spin--orbit and spin--exchange 
couplings of  coherent photoholes.
We validated 
our initial  prediction
of fs spin--orbit torque \cite{kapetanakis-2009}  
 by comparing our calculations 
to  existing 
 magneto--optical pump--probe
measurements,  
which monitor  the very early $\sim$100fs 
temporal regime following excitation 
with a single linearly--polarized pulse.
The most clear experimental signature 
is the observation of laser--induced fs magnetic hysteresis
and switching of 
the direction of 
out--of--plane 
femtosecond magnetization component 
with magnetic state. Such magnetic hysteresis 
is absent 
without  pump, while  
static planar Hall effect measurements
observe similar 
switchings  in the transverse 
component of the  
Hall magnetoresistivity.
Switching of the direction  of the  
laser--induced 
fs transverse  magnetization 
with magnetic state
cannot  arise from longitudinal 
 nonlinear optical effects and demagnetization/amplitude 
changes. 
The dependence on  magnetic state 
disappears  
with increasing perpendicular magnetic field, 
which  suppresses
the  magnetization re--orientation.
In this way we can separate experimentally 
longitudinal and transverse 
femtosecond magnetization changes. 
We discussed two theoretical results 
that may be useful for coherent control
of magnetic memory states and magnetic ringing via  
fs spin--orbit torque:  
(i) We showed that
femtosecond optical excitation 
can start, stop, and  restart 
 switching pathways
between the adiabatic  free energy 
magnetic states in any direction. 
Based on this, 
we gave an example of   sequences of   laser--pulse trains 
that can provide controlled access to
four different  magnetic states 
via  consequative 90$^{o}$
switchings,  
clockwise or counter--clockwise.
(ii) We demonstrated 
optical  control of magnon oscillations and switching rotations  
 and  suppression of  magnetic ringing  
at any time, long or short. 
For this we 
enhance spin--orbit torque
via pulse--shaping
and control  its direction 
via the frequency.

The model four--state  magnetic memory of Fig. \ref{Fig1}(b) 
allows for  verifiable 
tests to determine the feasibility  of our 
fs spin--orbit torque 
 proposal in 
realistic  materials. 
A full non--thermal control 
of   this four--state memory using our scheme  
requires the following:
(i) The competition between spin--orbit and magnetic exchange 
interaction breaks the symmetry 
while the laser electric field couples to the material, 
so   $e$--$h$ pair  excitations 
can be photoexcited with a 
finite spin.
There is 
no need to transfer angular--momentum 
from the photons 
(no circular polarization of the laser light), since 
spin--orbit coupling does not conserve spin. 
(ii) The direction, magnitude, and duration 
of the  non--thermal carrier 
spin  is coherently controlled by the 
optical field.
In particular, 
the direction of photoexcited spin 
is controlled by the laser frequency,
the magnetic state,
and the symmetry--breaking.
Importantly, its  
magnitude increases  with laser intensity
and ${\bf E}^2$, while 
its temporal profile 
follows that of the laser pulse
if  relaxation 
is faster.
Such  characteristics of  fs
spin--orbit 
torque  can 
distinguish it from 
adiabatic changes.
(iii) 
The  photoexcited carrier spin--pulses exert
fs spin--orbit--torques 
on the collective local spin 
and  move it  ``suddenly'',
 in a controllable direction
that depends on the magnetic state 
and the laser frequency.  
Important
is the pump--induced 
fs magnetic hysteresis for small magnetic fields,
absent in the static Kerr rotation angle
without pump. 
By coherently 
controlling the non--thermal population imbalance 
of 
exchange--split carrier bands 
with different spin--orbit interactions,
we can move the 
local spin via non--adiabatic interaction with 
mobile spins.  
(iv) Laser--pulse--shaping \cite{pulseshaping} 
and increased pump--fluence 
allow us to 
access optically the magnetic nonlinearities of the carrier 
free energy.
In this way,  
we may initiate or modify,
during fs timescales, deterministic 
switchings to 
any  available magnetic state.
(v) By 
using  control 
pulse--trains with appropriate frequencies, 
we suppress and  restart  switching rotations 
at intermediate magnetic states 
and suppress magnetic ringing 
after switchings complete. 
While coherent suppression 
of magnon oscillations 
is possible  
by taking advantage of the precession  phase,  
 here we mainly relied on controlling the  direction of 
 fs spin--orbit torque with respect to 
 the direction of magnetization rotation. In this way we 
 suppressed and enhanced 
both switching  and  ringing 
at long and   short times.

To control the 
entire four--state memory as in Fig. \ref{Fig9}, 
we had to use 
two  time--delayed laser--pulse--trains
with different frequencies
at each intermediate magnetic state.
The first excitation  suppresses the 
switching rotation/ringing in order to access the state,
while 
the second excitation 
restarts the process 
and moves the magnetization to the next magnetic state
in the desired direction. 
While such  control of the magnetization trajectory occurs 
on the 100fs timescale 
of coherent photoexcitation,  the 
initiated deterministic switchings
 complete on $\sim$100ps 
timescales, as determined by the free energy 
and  micromagnetic parameters. 
 In a massively--parallel memory, 
we can  control $n$ different bits
simultaneously  on the 
100fs timescale 
without waiting 
for each 
 switching 
 to   complete.
For large $n$, this   would ideally 
result in memory 
reading and writing at  $\sim$10THz 
 speeds.

Our  proposed
fs spin--orbit torque 
 mechanism  
may  be  relevant to 
different unexplored   spin--orbit coupled materials 
with coexisting mobile and local carriers,\cite{nagaev} 
for example
  topological insulators 
doped with magnetic impurities. \cite{TI,TI2} 
Important for  practical implementations 
and experimental proof of 
fs spin--orbit torque 
is to identify
materials where  
the quasi--thermal/adiabatic and 
non--thermal/non--adiabatic  contributions 
to the  magnetic anisotropy 
can be distinguished experimentally.
It is possible to separate these 
two 
based on their temporal profiles and 
their dependence 
on
photoexcitation
intensity, laser frequency, and external magnetic field. 
In (Ga,Mn)As, 
Fig. \ref{Fig3} 
shows photogeneration of a 
``sudden'' magnetization laser--induced re--orientation 
and fs magnetic hysteresis 
for magnetic field perpendicular to the 
sample plane.
Such magnetic field cants the ground state 
magnetization out of the plane, 
from $S_z$=0 ($B$=0) 
to $S_z$$\approx$$\pm S$ (large $B$). 
When $S_z$$\approx$0 in equilibrium, 
$\Delta S_z$(t) 
 measures transverse  magnetization re--orientation
and magnetic hysteresis correlated with in--plane 
switching,  
while when $S_z$$\approx$$S$ longitudinal 
changes dominate $\Delta S_z$(t)
and there is  no hysteresis.  
In this way, a perpendicular magnetic field 
can be used to elucidate the physical 
origin of the  fs  magneto--optical pump--probe 
signal  dynamics.
 Distinct thermal and non--thermal 
contributions to the ps magnetization 
trajectory
 were also observed experimentally at 
$\hbar \omega_p$$\sim$1.5eV. \cite{nemec-2013} 
They were separated   based
mainly on pump fluence 
dependence 
and by 
controlling the 
material's micromagnetic parameters. 
Qualitative  differences in the magnetization trajectory 
 were observed 
above $\sim$70 $\mu$J/cm$^{2}$
 pump 
fluence.
Below this,
the easy axis  
rotates smoothly 
inside the   plane, due to  
laser--induced 
 temperature  increase 
during $\sim$10ps timescales.
\cite{qi,nemec-2013} 
Above $\sim$70 $\mu$J/cm$^{2}$,
 a sub--picosecond ``sudden'' magnetization 
component is clearly observed. 
\cite{nemec-2013,qi}
Importantly, while 
the  precession frequency
$\gamma H_{FS}$ 
increases linearly with 
equilibrium temperature, 
it 
saturates
with pump fluence 
 above $\sim$70$\mu$J/cm$^{2}$,
even though 
the impulsive  out--of--plane
magnetization  tilt 
continues to increase. 
\cite{nemec-2013} 
In contrast, the pump--induced reflectivity
increases linearly
with pump intensity up to much higher 
fluences $\sim$150-200$\mu$J/cm$^{2}$, \cite{nemec-2013}
which indicates non--thermal photocarriers. 
At high intensities 
of 100's of $\mu$J/cm$^{2}$, 
fs demagnetization through dynamical polarization of
 longitudinal hole
spins dominates.  \cite{spinrelax}

In closing, we note that 
the discussed  concepts 
 are of 
more general applicability to condensed matter systems. 
The main idea 
is the possibility to  tailor 
order parameter 
dynamics 
 via optical coherent control 
of non--thermal carrier poulations, as well as via  
charge fluctuations and interactions 
driven while the optical field couples 
to the material.
The initial coherent excitation temporal regime may warrant 
more attention 
in various condensed matter systems. 
\cite{li-2013,Li-2}
An analogy can be drawn to the well--known coherent control of 
 femtosecond chemistry and 
photosynthetic dynamics, where the photoproducts of chemical and 
biochemical reactions can be influenced by 
creating coherent superpositions of molecular
 states.\cite{brumer} 
 Similarly, in condensed matter systems, laser--driven 
$e$--$h$ pairs (optical polarization) 
 can tailor 
 non--adiabatic ``initial conditions'' 
that drive  subsequent phase dynamics
governed by the free energy. 
An analogy can also be drawn 
to  parameter 
quenches studied in cold atomic gases. 
 ``Quasi--instantaneous'' quenches 
 drive  dynamics
that,  
in some cases such as  BCS superconductors,  
can be mapped to classical spin dynamics.
Coherent dynamics of superconducting order 
parameters 
are now beginning to be  studied 
in condensed matter systems, 
\cite{BCS,shimano}
and an analogy to the magnetic order parameter 
studied here is clear. 
Other  examples include
quantum femtosecond  magnetism
 in strongly--correlated manganites,
\cite{li-2013,Li-2}
photon--dressed Floquet states
in topological insulators, \cite{floquet}
or the existence of 
 non--equilibrium 
phases in charge--density--wave 
\cite{Porer} and other correlated systems. 
Femtosecond nonlinear optical  spectroscopy offers 
the time resolution needed to  
disentangle different 
 order parameters that 
are strongly coupled 
in the ground state, 
 based on their 
different dynamics after  ``sudden'' departure from 
equilibium. \cite{wang-nc,Porer}
Multi--pulse switching protocols
based on non--adiabatic quantum excitations
can  control 
non--equilibrium phase transitions,
by 
initiating  phase dynamics in a controllable way. 
\cite{li-2013,Li-2}

\section{Acknowledgements}

This work was supported by the
European Union's Seventh Framework Programme (FP7-REGPOT-2012-2013-1)
under grant agreement No. 316165, 
by 
the EU Social Fund and National resources 
through the THALES program NANOPHOS,
by the Greek Ministry of Education
ARISTEIA--APPOLO, 
  and by  the National Science Foundation Contract No. DMR-1055352.

\appendix 

\section{Fermi--Dirac/Adiabatic  versus Non--thermal/Non--adiabatic  
Magnetic Anisotropy}
\label{nonadia} 

In this appendix we discuss the two contributions
to laser--induced anisotropy, 
non--thermal  and quasi--thermal. 
The adiabatic/quasi--thermal  
contribution comes from relaxed 
Fermi--Dirac carriers.
The 
non--adiabatic contribution comes from the 
coherent/non--thermal photoexcited carriers, 
whose populations increase with intensity
during photoexcitation. 
In the initial stage, 
these non--thermal 
carriers  come from the 
continuum of 
$e$--$h$ excitations  
excited by the 
fs laser pulse, 
so they follow its temporal profile. 
At a second stage, 
they redistribute among the different ${\bf k}$ and 
band states while also
scattering with the Fermi sea carriers.

\subsection{Non--thermal/non--adiabatic  
magnetic anisotropy} 

We 
use
density matrix equations of motion and 
bandstructure to  
describe the  
femtosecond photoexcitation
of short--lived
 photohole spin--pulses driven by  
four competing effects:
(i) magnetic exchange interaction
between local and mobile spins,  
(ii) spin--orbit coupling of the 
mobile  carriers, 
(iii) coherent  nonlinear 
optical processes, and (iv) 
fast carrier relaxation. 
The interplay of these 
contributions  breaks the symmetry 
and excites
a controllable  fs  magnetic anisotropy field due to 
non--thermal photocarriers.  
The photoexcited spin, 
Eq.(\ref{h-spin}),  
is expressed in terms of the 
electronic density matrix, 
which resolves the different band and 
${\bf k}$--direction 
contributions.
Density matrix 
equations of motion were
derived for the time--dependent 
 Hamiltonian $H(t)$,  
Eq. (\ref{H}), 
with bandstructure treated within 
standard tight--binding and mean--field approximations. 
This  Hamiltonian has fast and slow
contributions. 
 Its adiabatic 
part $H_b({\bf S}_0)$, 
Eq. (\ref{Hb}),  
depends on the slowly varying 
(ps) spin  ${\bf S}_0$. 
The eigenstates of $H_b({\bf S}_0)$
   describe electronic bands
determined by periodic potential, 
spin--orbit, and adiabatic magnetic exchange
couplings. The latter, 
\begin{equation}
H_{pd}({\bf S}_0) 
 = \beta c \, {\bf S}_0 
 \cdot \hat{{\bf s}}_h,  
\label{DH}
\end{equation}
where 
 $\hat{{\bf s}}_{h}$
is the hole spin operator,  
leads to exchange--splitting of the  
 HH and LH semiconductor valence bands  determined  
by the  exchange  energy
$\Delta_{pd}$=$\beta c S$.
It also modifies the direction of photoexcited spin,  
by competing with the  
spin--orbit coupling of the mobile carriers
characterized by  
the energy splitting  $\Delta_{so}$  of the 
spin--orbit--split 
valence band of  the parent 
material (GaAs) at
 ${\bf k}$=0.
By adding to  the Hamiltonian 
carrier--carrier and carrier--phonon interactions,
 we can also 
treat 
relaxation, 
included 
here 
by introducing the 
non--thermal population relaxation time $T_1$
and the 
$e$--$h$ dephasing time $T_2$.

We describe  the band eigenstates 
of the adiabatic  electronic 
Hamiltonian $H_b({\bf S}_0)$ 
by using the 
semi--empirical tight--binding model 
that  reliably  describes  the 
GaAs bandstructure. \cite{vogl}
Compared to the 
standard  ${\bf k}\cdot{\bf p}$ 
effective mass approximation,
 this 
tight--binding approach 
allows us to  address states
with large momenta ${\bf k}$. 
Such  anisotropic and non--parabolic band states
 contribute  for 
laser frequencies 
away from the band--edge.
Following Ref.[\onlinecite{vogl}], we include the 
quasi--atomic 
spin--degenerate  orbitals 
3$s$, 3$p_x$, 3$p_y$, 3$p_z$, and 4$s$ 
of the two atoms per GaAs unit cell and use  the 
 tight--binding parameter values of the 
Slater--Koster sp$^3$s$^*$ 
model.
We add to this  description
of the parent material 
the mean--field coupling of the  Mn spin,  
Eq.(\ref{DH}), which 
modifies  spin--mixing in a non--perturbative way.
Similar to Ref.[\onlinecite{vogl}], 
we diagonalize the  Hamiltonian 
$H_b$=$H_b^c 
+ H_b^v$ 
to obtain the conduction 
($H_b^c$) and valence 
($H_b^v$)  bands
sketched in Fig. \ref{Fig1}(a):
\begin{eqnarray} 
 H_b({\bf S}_0) =\sum_{{\bf k} n} \varepsilon^c_{{\bf k} n}
\, \hat{e}^\dag_{{\bf k} n} 
\hat{e}_{{\bf k} n}  + \sum_{{\bf k} n} \varepsilon^v_{-{\bf k} n} \, 
\hat{h}^\dag_{-{\bf k} n} 
\hat{h}_{-{\bf k} n}.
\end{eqnarray} 
The eigenvalues 
 $\varepsilon^{c}_{{\bf k} n}({\bf S}_0)$
and  $\varepsilon^{v}_{-{\bf k} n}({\bf S}_0)$ 
describe  the 
conduction and valence  band energy dispersions.

While  ${\bf S}_0$ 
varies on a ps timescale  
much slower than the laser--induced electronic  fluctuations, 
the 
rapidly--varying (fs) 
part  of the Hamiltonian $H(t)$, 
 $\Delta H_{exch}(t)+H_{L}(t)$, 
drives ``sudden''  deviations from adiabaticity. 
$\Delta H_{exch}(t)$, 
Eq.(\ref{DHexch}), 
 describes 
non--adiabatic interactions 
of  photocarrier spins with the 
fs magnetization  $\Delta {\bf S}$(t)
 induced by 
fs spin--orbit torque.
$H_L(t)$ describes the optical field dipole coupling 
within the 
rotating wave 
approximation:
\begin{equation}
 H_{L}(t)= -
\sum_{n m {\bf k}} 
d_{n m {\bf k}}(t) \,  
\hat{e}^{\dag}_{{\bf k}m} \, \hat{h}^{\dag}_{{-\bf k}n}
+ h.c,
\label{HL} 
\end{equation}
where 
$d_{n m {\bf k}}(t)={\mu}_{n m {\bf k}} {\cal E}(t)$ 
is the Rabi energy,   ${\cal E}(t)$
is the pump electric field,
and  $\mu_{ n m {\bf k}}$ is the dipole transition 
matrix element
between the valence band $n$ 
and the conduction band $m$ at momentum 
${\bf k}$.  
These dipole  matrix elements
also depend on ${\bf S}_0$ and 
 are expressed in terms of the 
tight-binding parameters 
of $H_b({\bf k})$ 
as in  Ref.[\onlinecite{Lew}]: 
\begin{equation}
\mu_{nm{\bf k}}
=\frac{i}{\varepsilon_{m {\bf k} } - \varepsilon_{n {\bf k}}}
\langle n{\bf k}|\nabla_{\bf k} H_b({\bf k}) |m {\bf k} \rangle. 
\end{equation}

The density matrix  $\langle \hat{\rho} \rangle$
obeys  
the  equations of motion
\begin{equation}
i \hbar \frac{\partial \langle \hat{\rho} \rangle}{\partial t}=
\langle [\hat{\rho},H(t)] \rangle+ 
i \hbar \frac{\partial \langle \hat{\rho} \rangle}{\partial t}
|_{relax}.
\label{dm-eom} 
\end{equation}
The hole populations  and 
coherences  between 
 valence bands are given by the equation of motion 
 \begin{eqnarray} 
&& i  \hbar \, \partial_t  \langle \hat{h}^{\dag}_{-{\bf k}n} 
\hat{h}_{-{\bf k} n^{\prime}}\rangle
-
\left(\varepsilon^v_{{\bf k} n^{\prime}} 
- \varepsilon^v_{{\bf k} n} - i\Gamma^h_{nn^\prime} \right) 
\langle \hat{h}^{\dag}_{-{\bf k}n} 
\hat{h}_{-{\bf k} n^{\prime}}\rangle 
\nonumber \\
&& = \sum_{m}
d^*_{mn {\bf k}}(t) 
\, \langle \hat{h}_{-{\bf k}
n^{\prime}} \hat{e}_{{\bf k} m} 
\rangle
- \sum_{m}
d_{m n^{\prime} {\bf k}}(t)  
\, \langle  \hat{h}_{-{\bf k}
n} \hat{e}_{{\bf k} m}
 \rangle^*
\nonumber \\
&& 
+ 
\beta c 
\Delta {\bf S}
\sum_{l} 
\left[{\bf s}^h_{{\bf k} n^{\prime} l}
 \langle \hat{h}^{\dag}_{-{\bf k}n} 
\hat{h}_{-{\bf k} l}
\rangle -
 {\bf s}^{h*}_{{\bf k}n l} 
\langle \hat{h}^{\dag}_{-{\bf k} l} 
\hat{h}_{-{\bf k} n^{\prime}}\rangle\right],
 \label{dm-h} 
\end{eqnarray}
where
$n$=$n'$ describes the  non--thermal  populations and 
$n$$\ne$$n'$
the  coherent superpositions of 
different valence band states.
$\Gamma^h_{nn}$=$\hbar/T_1$ 
characterizes  the non--thermal population relaxation.  
$\Gamma^h_{nn^\prime}$
are 
the inter--valence--band dephasing 
rates, 
which are short and do not play an important role here. 
The first term on the rhs describes 
the photoexcitation of  hole  populations 
in  band states $(n,{\bf k})$ that 
depend on ${\bf S}_0$.
The second term is beyond a simple rate equation 
approximation
and describes the non--adiabatic changes 
in the hole states 
induced by their interaction
with the rapidly varying (fs)
 photoinduced
 magnetization
$\Delta {\bf S}$(t), 
Eq.(\ref{DHexch}).
Similarly,
 \begin{eqnarray} 
&& i  \hbar \, 
\partial_t 
\langle \hat{e}^{\dag}_{{\bf k}n} 
\hat{e}_{{\bf k} n^{\prime}}\rangle
-
\left(\varepsilon^c_{{\bf k} n^{\prime}} 
- \varepsilon^c_{{\bf k} n} - i \Gamma^e_{n n^\prime} \right) 
\langle \hat{e}^{\dag}_{{\bf k}n} 
\hat{e}_{{\bf k} n^{\prime}}\rangle= 
\nonumber \\
&&
 \sum_{m^{\prime}}
d^*_{n m^{\prime} {\bf k}} 
\langle \hat{h}_{-{\bf k}
m^{\prime}} \hat{e}_{{\bf k} n^{\prime}} 
\rangle 
- 
\sum_{m^{\prime}}
d_{n^{\prime}m^{\prime} {\bf k}} 
\langle  \hat{h}_{-{\bf k}
m^{\prime}} \hat{e}_{{\bf k}n }
 \rangle^*, 
 \label{dm-e} 
\end{eqnarray}
where the rates $\Gamma^e_{n n^\prime}$ characterize 
the  electron relaxation.

In the above equations of motion, 
the photoexcitation of the carrier populations 
and coherences 
is driven by the 
nonlinear $e$--$h$ optical  polarization 
$\langle \hat{h}_{-{\bf k} n} 
\hat{e}_{{\bf k} m} \rangle$ 
(off--diagonal density matrix element). 
This coherent amplitude  characterizes  the 
$e$--$h$  excitations 
driven by the  optical field, 
which here only exist during the laser pulse since their  
lifetime  $T_2$  (dephasing time) 
is short: 
 \begin{eqnarray} 
&&i \hbar \, \partial_{t} 
\langle \hat{h}_{-{\bf k} n} 
\hat{e}_{{\bf k} m} \rangle
- \left(\varepsilon^c_{{\bf k} m} 
+\varepsilon^v_{{\bf k} n} 
-i \hbar/T_2 \right) \langle \hat{h}_{-{\bf k} n} 
\hat{e}_{{\bf k} m} \rangle
\nonumber \\
&&
=
- d_{m n {\bf k}}(t) \, \left[1 - \langle \hat{h}^\dag_{-{\bf k}
 n} 
\hat{h}_{-{\bf k} n} \rangle 
- \langle \hat{e}^\dag_{{\bf k} m}
\hat{e}_{{\bf k} m} \rangle \right]
\nonumber \\
&&
+
\beta c \Delta {\bf S}(t) 
\cdot \sum_{n^\prime} 
{\bf s}^h_{{\bf k}nn^\prime} \
\langle \hat{h}_{-{\bf k} n^\prime} 
\hat{e}_{{\bf k} m} \rangle
\nonumber \\ && 
+ \sum_{n^\prime \ne n}
 d_{m n^\prime {\bf k}}(t) 
\, \langle \hat{h}^\dag_{-{\bf k} n^\prime} \hat{h}_{-{\bf k} n} \rangle 
\nonumber \\ && 
+  \sum_{m^\prime \ne m}d_{m^\prime n {\bf k}}(t) \, 
\langle \hat{e}^\dag_{{\bf k} m^\prime} \hat{e}_{{\bf k} m}
\rangle. 
\label{P}
\end{eqnarray} 
The nonlinear contributions to the 
above equation include 
 Phase Space Filling 
(first line),
transient  changes in the non--equilibrium 
hole states 
due to  the 
non--adiabatic magnetic exchange interaction $\Delta H_{exch}$(t) 
(second line), and   
coupling to 
$h$--$h$ (third line) and $e$--$e$  (fourth line) Raman  coherences. 
The  coupled Eqs. 
(\ref{dm-h}), (\ref{dm-e}), (\ref{P}),  
and (\ref{eom-Mn})
  describe   
photoexcitation of  non--thermal carriers
modified by the 
local spin  rotation.
They
were  
derived 
in Refs.[\onlinecite{chovan,kapetanakis-2009}]
using  Hartree--Fock factorization.
\cite{koch,rossi} 

\subsection{Adiabatic/Fermi--Dirac anisotropy} 

The equilibrium  mobile carriers can be described by
 Fermi--Dirac populations,  $f_{n {\bf k}}$,
of the eigenstates of the adiabatic Hamiltonian 
$H_b({\bf S}_0)$, which  determine 
the 
quasi--equilibrium  anisotropy field
${\bf H}_{FS}$, 
 Eq. (\ref{HFS}). 
\cite{kiriluk,bigot-chemphys,nemec-2013}
We simplify this 
thermal contribution by 
neglecting any
 laser--induced  changes in 
carrier temperature and 
chemical potential, 
which  add to our predicted  effects. 
A laser--induced thermal field 
$\Delta {\bf H}_{FS}$(t)  develops
indirectly from 
fs spin--orbit torque 
as the net spin of the hole Fermi sea bath 
 adjusts to the new   
non--equilibrium 
 direction of ${\bf S}$(t). \cite{chovan} 
As already seen
from calculations of magnetic anisotropy that assume a 
Fermi--Dirac distribution,
 \cite{nemec-2013,jung} 
the small ($\sim$$\mu$eV)
free energy differences with ${\bf S}$ result in 
anisotropy fields 
of the order of 10's of mT.  
The discrepancies  between theory and experiment 
seem to imply that non--equilibrium 
distributions that are broad in energy
are necessary in order to explain the magnitude of 
the observed effects.
 \cite{theory} 
Our time--domain calculation of 
laser--induced magnetic anisotropy 
driven by the photoexcited fs population and spin imbalance 
agrees with
experimental measurements.
However, we must still include the thermal Fermi sea 
anisotropy 
in order to describe the four--state magnetic memory. 
For this we  express 
the free
energy 
in the  experimentally--observed 
form dictated by symmetry, 
 \cite{jung,welp,dmwang}
also  obtained by expanding the theoretical 
expression: \cite{jung} 
\begin{equation}
\label{eq:Eh}
E_h({\bf S})
=
K_c(\hat{S}^2_x \hat{S}^2_y+ \hat{S}^2_x \hat{S}^2_z
+ \hat{S}^2_y \hat{S}^2_z)+K_{uz} \hat{S}^2_z-K_{u} \hat{S}_x \hat{S}_y,
\end{equation} 
where 
${\bf \hat{S}}$=${\bf S}/S$ is the unit vector 
that gives the instantaneous magnetization direction.
$K_c$ is the cubic anisotropy constant, 
$K_{uz}$ is the uniaxial constant, 
which includes  {\em both strain and shape
anisotropies}, and  $K_u$ describes an in--plane anisotropy
due to strain.  
We used  measured anisotropy parameter values 
\cite{dmwang}
$K_{c}$=0.0144meV, $K_{u}$=0.00252meV, 
and $K_{uz}$=0.072meV. We thus   obtain 
the thermal anisotropy field 
\begin{eqnarray}
&&  \gamma  {\bf H}_{FS}  = -  \frac{2K_c}{S}  {\bf \hat{S}} 
 + \frac{1}{S}(2K_{c}\hat{S}^{3}_{x}
+K_{u}{\hat S}_{y}, \nonumber \\
&&
2K_{c}{\hat S}_{y}^{3}+K_{u}{\hat S}_{x},
2K_{c}{\hat S}_{z}^{3}-2 K_{uz}{\hat S}_{z}).
\label{eq:nonli}
\end{eqnarray}
The above expression describes  the 
equilibrium magnetic nonlinearities 
of the 
realistic material.
By expressing ${\bf S}$ in terms of 
the polar angles $\phi$ and $\theta$, defined with respect to the 
crystallographic axes,
we obtain the easy axes
from the condition ${\bf S} \times {\bf H}_{FS}$=0, 
by solving the equations 
\begin{eqnarray}
&& 2 K_c \cos^3 \theta - (K_c+ K_{uz}) \cos \theta 
+ \frac{ B S}{2} =0
\label{anis-1} \\
&&
\sin 2\phi = \frac{K_u}{K_c\sin^2 \theta},
\label{anis-2} 
\end{eqnarray}
where we added the external magnetic field $B$
along the [001] direction. 
For $B$=0,
$\theta$=$\pi$/2 and 
Eq.(\ref{anis-2})  
 gives
the in--plane easy axes of  Fig. \ref{Fig1}(b).
For small $K_u$, 
these  magnetic states  
$X^+$, $X^-$,$Y^+$, and $Y^-$ 
are  tilted 
from the [100] and [010]    crystallographic directions
by few degrees inside the plane. \cite{qi,welp} 
As can be seen from 
Eq.(\ref{anis-1}),
the $B$--field  along [001] cants the 
 easy axes out of the plane.
 In this case, 
 $\theta$$\ne$$\pi/2$ and 
 Eq.(\ref{anis-2}) 
shows a simultaneous rotation inside the plane.
The above out--of--plane easy axis component,  
measured by the 
static Kerr rotation angle $\theta_K(B)$,
\cite{wang-2009,spinrelax} 
 varies 
smoothly with magnetic field. 
On the other hand, Eq.(\ref{anis-2}) gives two different 
values for 
$\phi$ ($X$ and $Y$), which can switch 
due to either $B$--field changes (as seen in the 
transverse Hall magnetoresistivity \cite{wang-2009})
or laser--induced 
fs spin--orbit torque (as predicted here).

\section{ Band Continuum of Electronic States} 
\label{cont} 

The average hole spin ${\bf s}_h$(t), Eq.(\ref{h-spin}), 
that triggers the fs magnetization dynamics  here 
has contributions 
${\bf s}_{{\bf k} n}^h$(t) from an anisotropic 
  continuum of 
photoexcited non--parabolic 
 band states.
At $\hbar \omega_p$$\sim$1.5eV, this  continuum 
includes disordered--induced states below the bandgap 
of the pure semiconductor. \cite{theory} 
At $\hbar \omega_p$$\sim$3.1eV, 
photoexcitation of such impurity band/defect states is small, 
while the almost parallel conduction and 
valence bands 
lead to 
 excitation of a wide range 
of  ${\bf k}$ states.
Integration over the Brillouin zone momenta, as in  
Eq.(\ref{h-spin}), 
presents a well--known challenge 
for calculating magnetic anisotropies and other
 properties of real materials.
\cite{anis} 
To simplify the problem,  one often 
calculates the  quantities of interest 
at  select  ${\bf k}$--points 
and replaces the integral by a weighted sum 
over these ``special points'' (special point approximation). 
\cite{anis}
In our previous work, \cite{kapetanakis-2009} 
we considered 
 eight  special  ${\bf k}$--points ($\Lambda$--point \cite{burch})
along  \{111\}.  
While this approximation takes into account
 the general features of the 
anisotropic states, 
it misses important  details, 
such as strong band non--parabolicity, density of states, 
and photoexcited carrier densities.
To compare with the photocarrier densities 
in the experiment and to address issues such as 
the frequency dependence 
of the photoexcited spins,  
we must include continua of 
band states 
in our calculation. 
Here we  
integrate over the 
 band--momenta
along 
the eight \{111\} 
 symmetry lines by
using the  ``special lines approximation'' 
discussed in 
Ref.[\onlinecite{enders}].
At
 $\hbar \omega_p$$\approx$3.1eV,
we approximate the 
three--dimensional momentum integral 
by a sum of one--dimensional integrals along the eight 
${\bf k}$ directions
populated by 
photoexcited carriers. 
This simple approximation includes the anisotropic, non--parabolic 
band continua.\cite{enders} 

Following 
Ref.[\onlinecite{enders}], 
we first  express 
\begin{eqnarray}
&& \frac{1}{V}\sum_{\bf k} 
\Delta {\bf s}^h_{{\bf k}}
=\frac{1}{(2 \pi)^3} \  \int_{BZ} \ \Delta {\bf s}^h_{{\bf k}} \   d{\bf k} \nonumber \\
&&
= \int \frac{d \Omega}{4 \pi} \ \left[
\frac{1 }{(2 \pi)^3} \int_{0}^{k_{BZ}} 4 \pi k^2 dk 
\Delta {\bf s}^h_{{\bf k}} \right],
\end{eqnarray} 
where $k_{BZ}$ is the Brillouin zone boundary 
and $d \Omega$ 
is the angular integral. 
To calculate the above angular--average, 
we use the special lines approximation \cite{enders} 
\begin{equation} 
\int \frac{d \Omega}{4 \pi} \ 
\Delta {\bf s}^h_{{\bf k}}
= \sum_{\alpha}  w_\alpha \,
\Delta {\bf s}^h_{k \alpha},
\end{equation} 
where $\alpha$  runs over the dominant 
symmetry 
directions, 
$k$ is the wavevector amplitude, 
 and $w_\alpha$ are weight factors. 
For $\hbar \omega_p \sim$3.1eV, the 
dominant contribution comes from the eight 
\{111\} symmetry directions, so we approximate 
\begin{eqnarray}  
\frac{1}{V}\sum_{\bf k} 
\Delta {\bf s}^h_{{\bf k}}
= 
\frac{1}{(2 \pi)^3} 
\sum_{\alpha=\{111\}}  w_\alpha 
\int_{0}^{k_{BZ}} 4 \pi k^2  \,
\Delta {\bf s}^h_{k \alpha} \ dk. 
\end{eqnarray} 
Instead of  
eight discrete ${\bf k}$--point populations  
as in Ref.[\onlinecite{kapetanakis-2009}], 
here we consider continuum distributions along the 
eight  one--dimensional ${\bf k}$--lines. 
While the estimation of optimum weight factors $w_\alpha$ 
is beyond the scope 
of this paper, \cite{anis} 
the order of magnitude of the predicted 
effects is not sensitive to their 
precise value. 
We fix  $w_\alpha$=$w$ 
by reproducing the net 
photohole density $n$ 
 at one experimentally--measured intensity: 
\begin{eqnarray} 
&& n=\frac{1}{V}\sum_{\bf k} \sum_n 
\Delta \langle \hat{h}^{\dag}_{-{\bf k} n} \,
\hat{h}_{-{\bf k} n} \rangle \nonumber \\ && 
= 
\frac{w}{(2 \pi)^3} 
\sum_{\beta=\{111\}} 
\int_{0}^{k_{BZ}} 4 \pi k^2  \,
\Delta \langle \hat{h}^{\dag}_{k \beta n} \, 
\hat{h}_{k \beta n} \rangle. 
\end{eqnarray} 
For the results of Fig. \ref{Fig3}, 
the  photocarrier density
 $n$$\sim$6$\times$10$^{18}$/cm$^3$
  for pump fluence 
$\sim$7 $\mu J/cm^{2}$
gives $w$$\sim$1/15. 
The same order of magnitude of $n$ is obtained, however,  for all 
other reasonable  values of $w$. \cite{enders}
We  then used this weight factor 
 for all other laser intensities.

\end{document}